\documentclass[ceqn,10pt]{wlscirep}
\usepackage[utf8]{inputenc}
\usepackage[T1]{fontenc}
\usepackage{xcolor}
\usepackage[table]{xcolor}
\usepackage{colortbl}
\usepackage{booktabs}
\usepackage{pdflscape}
\usepackage{multirow}
\usepackage{makecell}
\usepackage{cleveref}
\usepackage{subfig}
\usepackage{authblk}
\usepackage{amssymb}
\usepackage{amsmath} 
\usepackage{amsfonts}
\usepackage{latexsym}
\usepackage{amsthm} 
\usepackage{eucal}
\usepackage{wrapfig}
\usepackage{microtype}
\usepackage{tikz}
\usepackage{pgfplots}
\usepackage{caption}
\usepackage[subfigure]{tocloft}
\usepackage{enumerate}
\usepackage[overload]{empheq}
\usepackage{mathtools}
\usepackage{cases} 
\usepackage{url}
\usepackage{hyperref, xr-hyper}    
\usepackage{tabularx} 
\usepackage{longtable}
\usepackage{array}     

\crefname{figure}{Fig.}{Figs.}
\Crefname{figure}{Figure}{Figures}
\crefname{table}{Table}{Tables}
\Crefname{table}{Table}{Tables}

\addto\captionsenglish{
	\renewcommand{\contentsname}%
	{Supplemental Information:\\ The networks of ingredient combination in cuisines around the world}}

\newcommand{\listequationsname}{Supplementary Equations}
\newlistof{myequations}{equ}{\listequationsname}

\addto\captionsenglish{
	}
\def\l@section{\@tocline{1}{0,2pt}{2pc}{12mm}{\ \ }} 

\newcommand{\startsupplement}{
	\setcounter{figure}{0}
	\setcounter{table}{0}
	\setcounter{equation}{0}
	\renewcommand{\thetable}{S\arabic{table}}
	\renewcommand{\theequation}{S\arabic{equation}}
	\renewcommand{\thefigure}{S\arabic{figure}}
	\renewcommand{\cftfigpresnum}{Fig.~}
	\renewcommand{\cfttabpresnum}{Tab.~}
	\setlength{\cftfignumwidth}{5.5em}
	\setlength{\cftfigindent}{0em}
	\setlength{\cfttabnumwidth}{5.5em}
	\setlength{\cfttabindent}{0em}
}
\newcommand\blfootnote[1]{
	\begingroup
	\renewcommand\thefootnote{}\footnote{#1}
	\addtocounter{footnote}{-1}
	\endgroup
}

\title{The networks of ingredient combinations as culinary fingerprints of world cuisines }

\author[1,+]{Claudio Caprioli}
\author[2,3,+]{Saumitra Kulkarni}
\author[4]{Federico Battiston}
\author[5,6,$\dagger$,*]{Iacopo Iacopini}
\author[7,8,$\dagger$,*]{Andrea Santoro}
\author[1,9,10,$\dagger$]{Vito Latora}
\affil[1]{Department of Physics and Astronomy, University of Catania, 95125 Catania, Italy}
\affil[2]{Network Science Institute, Northeastern University, Boston, MA 02115, USA}
\affil[3]{Department of Scientific Computing, Modeling \& Simulation, Savitribai Phule Pune University, 411007 Pune, India}
\affil[4]{Department of Network and Data Science, Central European University, 1100 Vienna, Austria}
\affil[5]{Network Science Institute, Northeastern University London, E1W 1LP London, United Kingdom}
\affil[6]{Department of Physics, Northeastern University, Boston, MA 02115, USA}
\affil[7]{CENTAI Institute, Turin, Italy}
\affil[8]{Neuro-X Institute, {\'E}cole Polytechnique F{\'e}d{\'e}rale de Lausanne, 1202 Geneva, Switzerland}
\affil[9]{School of Mathematical Sciences, Queen Mary University of London, E1 4NS London, United Kingdom}
\affil[10]{Complexity Science Hub, 1080  Vienna, Austria}
\affil[+]{these authors contributed equally to this work}
\affil[$\dagger$]{these authors contributed equally to this work}

\begin{abstract} 
	Investigating how different ingredients are combined in popular dishes is crucial to uncover the principles behind food preferences. Here, we use data from public food repositories and network analysis to characterize and compare worldwide cuisines. Ingredients are first grouped into broader types, and each cuisine is then represented as a network in which nodes correspond to ingredient types and weighted links describe how frequently pairs of types co-occur in recipes. Cuisines differ not only in the popularity of ingredient types and range of recipe sizes, but also in the structural organization of ingredient-type combinations. By analyzing these networks, we uncover distinctive patterns of type associations that serve as culinary fingerprints. For example, European cuisines typically distribute ingredients across different types, whereas certain Asian and South American traditions emphasize one dominant type, such as vegetables or spices. The essence of these patterns is well captured by the networks' maximum spanning trees, which offer a simplified yet representative backbone for each cuisine. We demonstrate that both these full and simplified network representations enable machine learning models to identify cuisines from subsets of recipes with very high accuracy. Networks of ingredient combinations also cluster global cuisines into meaningful geo-cultural groups, reflecting shared patterns in culinary traditions. More broadly, our study offers novel insights into the structure of world cuisines, enabling data-driven approaches to their characterization, cross-cultural comparison, and potential adaptation.
\end{abstract}

\begin{document}
	
	\flushbottom
	\maketitle
	\blfootnote{$^*$corresponding authors: \href{iacopo.iacopini@nulondon.ac.uk}{iacopo.iacopini@nulondon.ac.uk}, \href{andrea.santoro@centai.eu}{andrea.santoro@centai.eu}}
	\thispagestyle{empty}
	
	\section*{Introduction}
	The act of transforming raw ingredients into palatable sustenance, known as cooking, predates the appearance of \textit{Homo sapiens}~\cite{Bottèro_1987, Nadeau2009, Zohar2022}. From the earliest civilizations, recipes have served as templates, guiding individuals in the preparation of meals using specific techniques and ingredients~\cite{10.1093/oxfordhb/9780199729937.001.0001}. Over millennia, culinary traditions have undergone constant refinement and expansion, reflecting the dynamic interplay of locally available ingredients with cultural influences, trade, technological advancements, and changing palates. As a result of this long and intertwined process, specific recipes have become the very backbone of diverse culinary traditions across the globe~\cite{diamond1997guns, civitello2004cuisine}.
	While traditional recipe books have long served as repositories of culinary knowledge, the advent of computational approaches has opened new avenues for understanding recipes as algorithms~\cite{goel2022computational}. Researchers have recently started leveraging vast data sets of recipes to reveal hidden patterns and extract valuable insights into culinary practices~\cite{ratatouille_2022, Yamakata_2013, Mori2012AML, Wu_2021}. Tools coming from natural language processing~\cite{Khurana2023} and machine learning~\cite{bishop2013pattern} have revolutionized our ability to analyze and interpret the complexities of recipe composition and preparation methods~\cite{marin2019learning}. Applications range from extracting the list of ingredients from meal images~\cite{salvador2019inverse, channam2021recipe}, 
	to examining nutrient concentration in food~\cite{menichetti2022nutrient}, predicting food processing levels~\cite{menichetti2023machine} and investigating the impact of certain diets on health~\cite{aielloLargescaleHighresolutionAnalysis2019,wang2023nutritional}. 
	
	The essence of recipes cannot be separated from the ingredients, which as in a symphony composed of individual notes, come up together to create an unexpected harmonic consonance, each contributing its unique flavor and texture to the final dish. It is precisely this combination of ingredients ---and the different techniques used to cook them--- that would ultimately determine whether the resulting product would be successful or not. At the heart of culinary creativity lies a quest to explore and discover novel ingredient combinations — a journey in a virtually unbounded space of pairings, mixtures, and innovations that has only been partially charted. However, not all ingredient combinations are successful, as evidenced by the considerable gap between the potential number of recipes and those recorded throughout human history~\cite{Kinouchi_2008}. This discrepancy suggests that culinary creativity often builds on top of known flavors, balancing the search for novelty with the use of established, successful combinations~\cite{kauffman1996investigations, tria2014dynamics, iacopini2018network}. 
	
	Complex networks ~\cite{newman2006structure, complex_networks_latora}
	are particularly suited to describe the way in which different ingredients are combined in recipes, and also to capture recurring patterns ~\cite{Ahnert2013, bellingeri_2025}. 
	So far, networks have been applied to describe ingredient combinations with a particular focus on their chemical composition. In particular, bipartite networks~\cite{guillaume2004bipartite, guillaume2006bipartite, torres2021and} have been instrumental in studying the {\it food pairing hypothesis}, a famous culinary conjecture introduced by Chef Heston Blumenthal~\cite{blumenthal2008big}, stating that the compatibility of ingredients in recipes strongly depends upon the amount of shared flavor compounds~\cite{ahnFlavorNetworkPrinciples2011, Jain2015SpicesFT, Jain2015-yz, 8402036}. Recent findings however suggest that food pairing is not a universal principle, but cuisines worldwide differ in the role chemical affinity plays in recipe formation~\cite{Simas2017}.
	The growing interest in the chemical composition of food has spurred broader explorations into the implications of food consumption on health. Network-based studies have significantly advanced biomedical knowledge by unraveling the intricate interplay of nutritional value, chemical profiles, and their impact on diseases~\cite{aielloLargescaleHighresolutionAnalysis2019, Barabási2020, kimUncoveringNutritionalLandscape2015, menichetti2022nutrient, Cenikj2023}. 
	It is now well understood that exploring the chemical complexity of food is an essential requirement for the study of human health, inextricably linked to the diet of individuals~\cite{menichetti2025food_review}. However, while analyses focused on chemical composition are pivotal in clarifying the relationship between diet and health, and at the same developing targeted therapies and precision nutrition, they often fail to capture the taste profiles of ingredients. Assortative principles based on food compounds lack generality and apply only to specific geographic cuisines. Moreover, these analyses typically overlook how culinary traditions combine ingredients in distinct and culturally meaningful ways. While reductionism is often a driving force in scientific inquiry, its application to gastronomy hides the risk of discarding valuable information encoded in culinary practices, knowledge long recognized and applied in kitchens around the world. A crucial observation is that ingredients often serve similar functional roles within recipes, regardless of their chemical makeup, such as providing umami, acidity, or aromatic complexity. For instance, soy sauce in East Asian cuisine and fish sauce in Southeast Asian cuisine both function as salty umami enhancers. Similarly, very different types of fat sources, like oil, butter, or lard, are often used as non-stick agents and heat conductors when cooking other ingredients in a pan. This suggests that studying ingredients based on their functional associations, rather than in isolation, could reveal hidden patterns in culinary practices.
	
	In this work, we address these limitations by categorizing ingredients into broader ``food types'' or ``ingredient categories''. Drawing upon a recipe dataset consistently covering geographic areas from all the five continents~\cite{8402036}, 
	we employ a network-based approach to characterize and compare cuisines from all over the world. 
	Each regional cuisine is turned into a network of ingredient combinations, 
	whose nodes represent food types, while links describe the ways food types are combined in recipes. The analysis of such networks allows to identify unique food profiles and food co-occurrence patterns associated with each cuisine, 
	highlighting with unprecedented precision commonalities and differences in gastronomic preferences across geographic regions.
	
	\section*{Results}
	
	\begin{figure}[t]
		\centering
		\includegraphics[width=0.5\columnwidth]{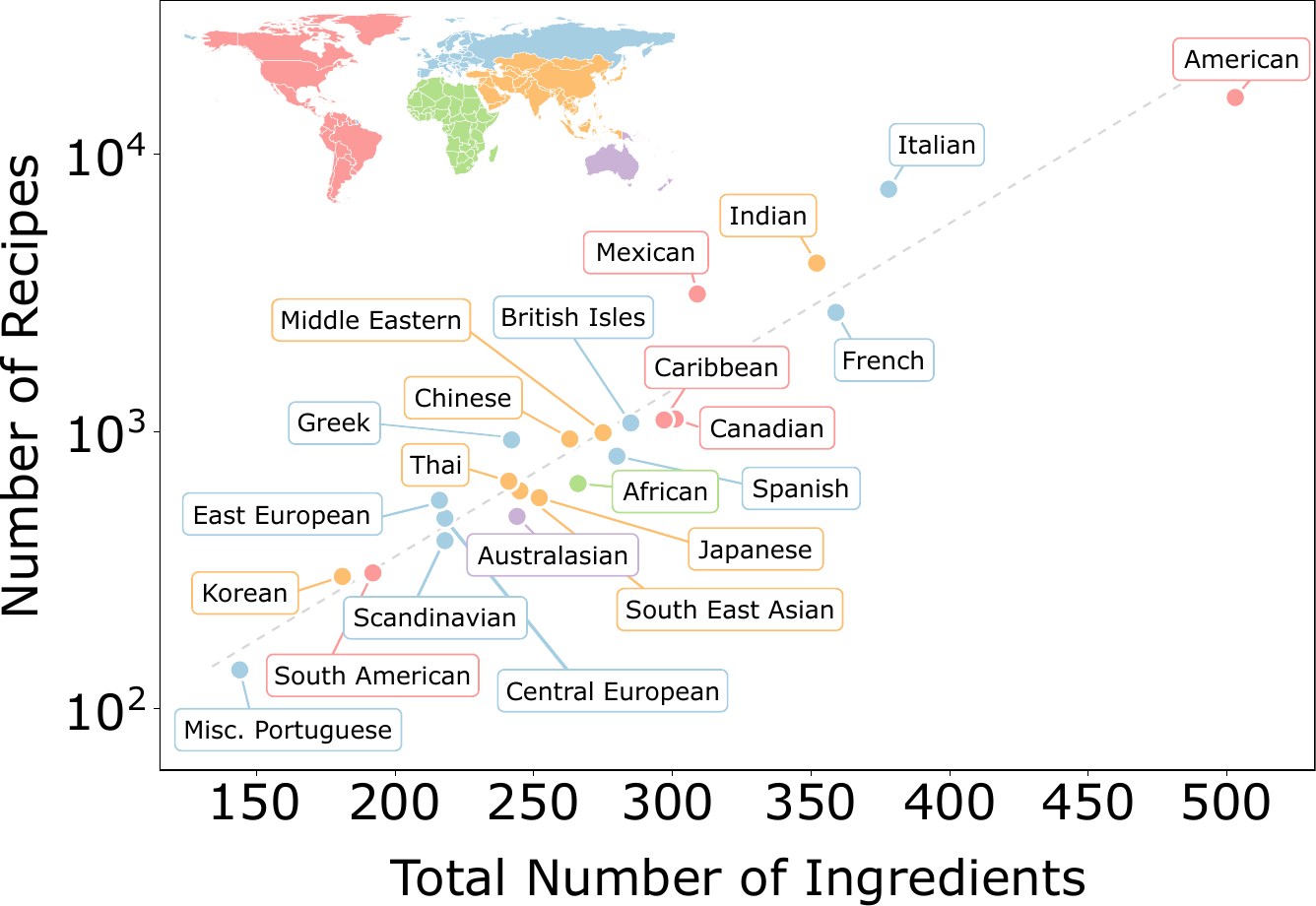}
		\caption{	\textbf{Scaling of recipes vs total ingredients across world cuisines.} Relationship between the number of recipes and the total number of individual ingredients across various cuisines. Each data point represents a cuisine and is color-coded according to the continent affiliation, as illustrated in the world map reported in the top-left corner. A lin-log scale has been used in the plot, and the reported best-fit line highlights the exponential growth of the number of recipes with the total number of ingredients.}
		\label{fig:recipe_ingredients_scaling}
	\end{figure}
	
	\subsection*{Statistical descriptors of world cuisines}
	We study a data set of recipes (dishes) covering 23 cuisines around the world, for a total of $R=45,661$ recipes and $I=604$ unique ingredients (see Methods for a detailed description of the data set). This number results from a preprocessing stage aimed at standardizing the ingredient set and reducing dataset complexity, which consolidated the original 695 raw ingredient entries into 604 canonical forms. That is, ingredients with similar culinary roles or origins were grouped together, for instance, merging varieties (e.g., all pasta types as pasta), preparation forms (e.g., dried and fresh peas as peas), and subtypes (e.g., cherry and plum tomatoes as tomatoes). Additionally, ingredients produced through similar transformation processes were mapped to unified labels (e.g., all types of cheese, bread, and wine), and animal parts were aggregated by species (e.g., all chicken parts as chicken). Full details on the preprocessing are provided in the Methods and Supplementary Note 1. Each cuisine comes with its own set of recipes, where each recipe is a different combination of ingredients. Starting from the individual ingredients, we investigate how the number of ingredients and the number of recipes vary across the different cuisines. In \cref{fig:recipe_ingredients_scaling}, we report the total number of recipes as a function of the total number of ingredients in a lin-log scale. The distribution of distinct ingredients varies significantly among cuisines, ranging from the 144 of the Portuguese cuisine to the 503 of the American one\footnote{Acknowledging the ambiguous use of this term, we adopt the standard terminology where American refers to dishes prepared in the United States only (\url{https://en.wikipedia.org/wiki/American_cuisine})}. Notably, the combination of ingredients in recipes differs widely among cuisines, with cuisine sizes spanning three orders of magnitude. The smallest and largest recipe data sets are again represented by Portuguese and American cuisine, with 138 and 16,056 recipes respectively. This heterogeneity suggests a substantial difference in the  way ingredients are combined and possibly also in their frequency of use in recipes, consistently with previous findings on different data sets~\cite{Kinouchi_2008,ahnFlavorNetworkPrinciples2011}. The exponential growth of recipes with the number of ingredients (best-fit exponent $\gamma$ = 0.1, $R^2$=0.77) allows a broad classification of cuisines into two groups ---based on their position relative to the best-fitting curve. Cuisines above this curve, such as the Italian cuisine, feature more recipes than expected from their rough ingredient count (e.g., 378 ingredients forming 7,493 recipes, surpassing the expected 4,288). Conversely, cuisines below the curve indicate fewer recipes than those expected based on the number of ingredients employed. One example is the Portuguese cuisine, which has 144 ingredients but only 138 recipes. 
	
	\begin{figure*}[t!]
		\centering
		\includegraphics[width=1\textwidth]{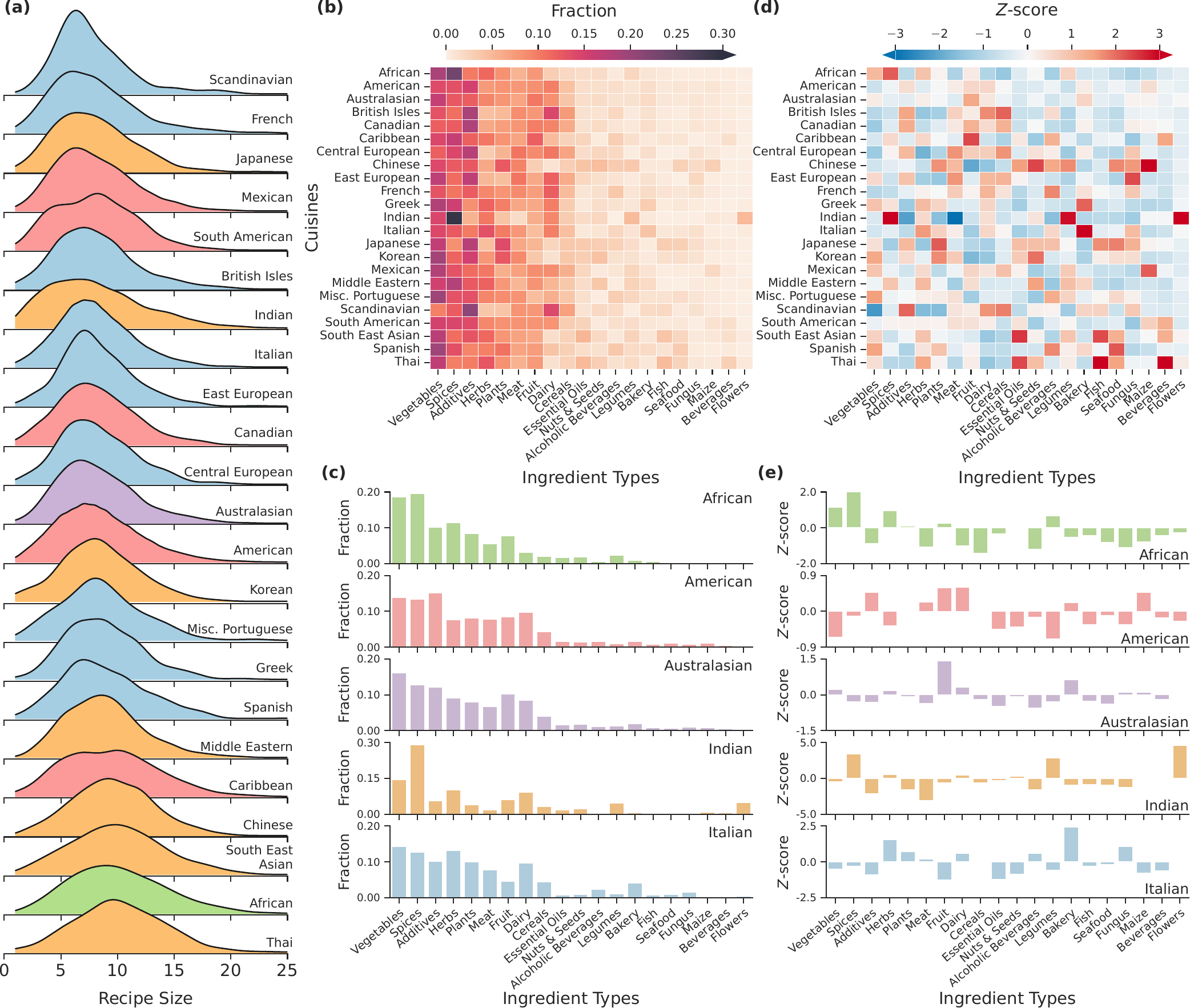}
		\caption{	\textbf{Recipe sizes and ingredient popularity.} \textbf{(a)} Ridge plot illustrating the distribution of recipe sizes across cuisines from different continents. Each distribution, sorted by average, corresponds to a cuisine, and is colored according to the continent affiliation (as per \cref{fig:recipe_ingredients_scaling}). \textbf{(b)} Heatmap depicting the ingredient type popularity profiles of cuisines. Each row corresponds to a cuisine, and each column to an ingredient type. The color intensity represents the popularity of an ingredient type in a given cuisine. \textbf{(c)} Bar chart representation of ingredient type popularity profiles derived from panel (b) for five specific cuisines, highlighting the variations in ingredient preferences. \textbf{(d)} Heatmap illustrating the z-scores of the ingredient type popularity profiles, providing insights into their relative significance across cuisines. Dark-red and dark-blue regions indicate respectively higher and lower z-scores. \textbf{(e)} Bar chart representation of the z-scores derived from panel (d) for the same five cuisines, revealing the relative significance of ingredient types.}
		\label{fig:recipe_sizes_ingredient_popularity}
	\end{figure*}

	The scaling of Fig.~\ref{fig:recipe_ingredients_scaling} reveals differences between cuisines at a macroscopic level, hiding preferences in the use of ingredients and on the way these are combined in recipes. To gain insights into these aspects, we examine the distribution of recipe sizes across cuisines. 
	Figure~\ref{fig:recipe_sizes_ingredient_popularity}~(a) shows the probability distribution $P_r (n_r=n )$ that a randomly chosen recipe of a given cuisine has $n$ ingredients.  
	Each distribution represents a specific cuisine and is color-coded according to the cuisine's continent affiliation (as depicted in \cref{fig:recipe_ingredients_scaling}).  
	Cuisines are ordered from top to bottom based on their average recipe size. 
	This analysis reveals two distinct tendencies: European (especially Continental and Southern European) cuisines tend to favor on average smaller recipe sizes, while Asian cuisines exhibit a preference for larger recipe sizes.
	Accordingly, the recipe with the largest number of ingredients in the data set is \texttt{Vegetable Korma} — a traditional Indian dish made with mixed vegetables, spices, and dairy products — which uses a staggering amount of 31 ingredients (see Supplementary Table~S1 where we report the largest recipe of each cuisine, with the respective ingredients list).

	A step further in highlighting the characteristics of cuisines is to focus on which specific ingredients are utilized and mixed in a recipe. 
	In order to do this, we consider the division of ingredients into macro-categories or ``ingredient 
	types''.  
	In our data set, ingredients are classified into 20 different types, namely:  
	Additive, 
	Bakery, 
	Beverages,
	Alcoholic Beverages,
	Cereals, 
	Dairy,
	Essential Oils, 
	Fish,
	Flowers,
	Fruit,
	Fungi,
	Herbs,
	Legumes,
	Maize,
	Meat,
	Nuts \& Seeds,
	Plants,
	Seafood,
	Spices,
	Vegetables. 
	The rationale behind this classification, inherited from the original dataset~\cite{culinarydb_webpage}, is described in detail in the Methods section.
	We define the popularity of an ingredient type within a given cuisine as the fraction of ingredients of that type that constitute the recipes of that cuisine. 
	Ingredient popularity enables us to shed light on how different cuisines utilize various ingredient categories within their recipes. 
	To achieve this, in \cref{fig:recipe_sizes_ingredient_popularity}~(b) we report a heatmap encoding the ingredient popularity profiles. Rows represent individual cuisines and columns represent specific ingredient types. To compute popularity profiles, we count the total number of occurrences of all ingredients belonging to a specific type across all recipes within a given cuisine. To account for potential biases arising from varying numbers of recipes or their lengths, we normalize these counts by dividing them by the sum of all recipe sizes within that cuisine.
	The heatmap illustrates a general pattern of ingredient usage across cuisines, with darker colors indicating higher popularity of certain ingredient types (on average, 9 types out of 20 are more popular than others).
	This common trend in ingredient usage is characterized by darker regions concentrated towards the left-hand side of the heatmap, indicating a higher popularity of certain ingredient types; as one moves towards the right-hand side, the popularity of ingredient types decreases and colors gradually lighten.
	However, despite this overall trend, individual cuisines exhibit distinct preferences. This becomes evident when individually examining the distribution of popularity profiles, as done for five specific cuisines in \cref{fig:recipe_sizes_ingredient_popularity}~(c) (see Fig.~S1 for the popularity profiles of the remaining cuisines).
	Furthermore, to emphasize differences among cuisines, we report in the heatmap of \cref{fig:recipe_sizes_ingredient_popularity}~(d) the $z$-scores of the ingredient type popularity.
	Here, regions ranging from light-red to dark red in the heatmap,  denote progressively higher positive $z$-scores, indicating that a cuisine tends to use certain ingredient types more frequently than others.
	Conversely, regions with light to dark blue colors represent increasingly negative $z$-scores, indicating below-average usage of certain ingredient types. For the sake of completeness, we report in \cref{fig:recipe_sizes_ingredient_popularity}~(e) the z-scores distributions for five specific cuisines (see Fig.~S2 for the z-scores distributions of the remaining cuisines).
	Notably, \cref{fig:recipe_sizes_ingredient_popularity}~(d) and (e)
	reveal some distinctive traits of cuisines, which can be attributed to geographical, cultural or religious reasons. For example, India stands out as the cuisine that uses meat the least, with a high negative $z$-score, which reflects cultural restrictions on meat consumption.
	In contrast, Spices, Legumes and Flowers are used in Indian recipes with much higher frequency compared to other cuisines, aligning with the popular consensus of Indian cuisine as predominantly vegetarian, with extensive use of Flowers-based cooking oils and aromatic Spices.
	Similarly, the impact of geography on ingredient usage is exemplified by Scandinavian cuisine, which shows significantly lower usage of Vegetables, Herbs, and Plants compared to most other cuisines. 
	The harsh climate of Scandinavian countries creates unfavorable conditions for the cultivation of most Vegetables, resulting in a distinct culinary approach that relies on other ingredients. 
	
	\subsection*{The networks of ingredient combinations}
	
	A cuisine is not solely characterized by the quantity and type of used ingredients, but also by the way in which the ingredients are combined.
	To describe the intricate patterns of ingredient combinations within recipes and better capture how such relationships vary across different cuisines, we introduce the network of ingredient combination. 
	Recipe data sets can be represented as a bipartite graph, where ingredients and recipes are the two disjoint sets of nodes (see Methods for details). 
	The projection of such networks on the set of ingredients (see Methods) allows us to obtain unipartite ingredient-ingredient graphs which provide summarised information on how ingredients are combined in recipes, while sensibly reducing the number of nodes of the original graph. 
	However, the projection graphs present two major obstacles: (\textit{i}) the size reduction from bipartite to unipartite is not sufficient, leaving the resulting graphs overly complex for interpretation; (\textit{ii}) the comparison between graphs might be inherently biased due to variations in graph size and ingredient sets across cuisines.
	To address these issues, we consider a coarse-grained representation of the ingredient-ingredient graphs, hereafter referred to as \textit{ingredient type-ingredient type 
		graphs}, or simply 
	\textit{type-type 
		graphs}. Nodes in such graphs correspond to the 20 distinct ingredient types in our data set.
	An edge connects ingredient types $t$ and $t'$ if there is at least one recipe of the cuisine containing at least a pair of ingredients respectively of type $t$ and $t'$. 
	The edge weight reflects the frequency of such pairings across the entire data set. 
	For example, the "apple strudel" recipe of Central European cuisine, comprising a total of 4 ingredients (apples, raisins, butter, puff pastry), ensures the existence of the 4 following edges: Fruit-Fruit, Fruit-Dairy, Fruit-Bakery, Bakery-Dairy. In particular, such recipe contributes with a weight of 2 to the two edges Fruit-Dairy and Fruit-Bakery in the network, reflecting the fact that butter and puff pastry are paired with two different fruits. 
	The type-type graphs encode the main features of how ingredients are combined, without the complexity of a description at the level of individual ingredients. Moreover, it offers a uniform representation of culinary traditions across different cuisines. Being all made by the same number of nodes, our networks allow for direct comparisons of worldwide cuisines, revealing commonalities and discrepancies in ingredient interactions.
	
		\begin{figure*}[t]
		\centering
		\includegraphics[width=\textwidth]{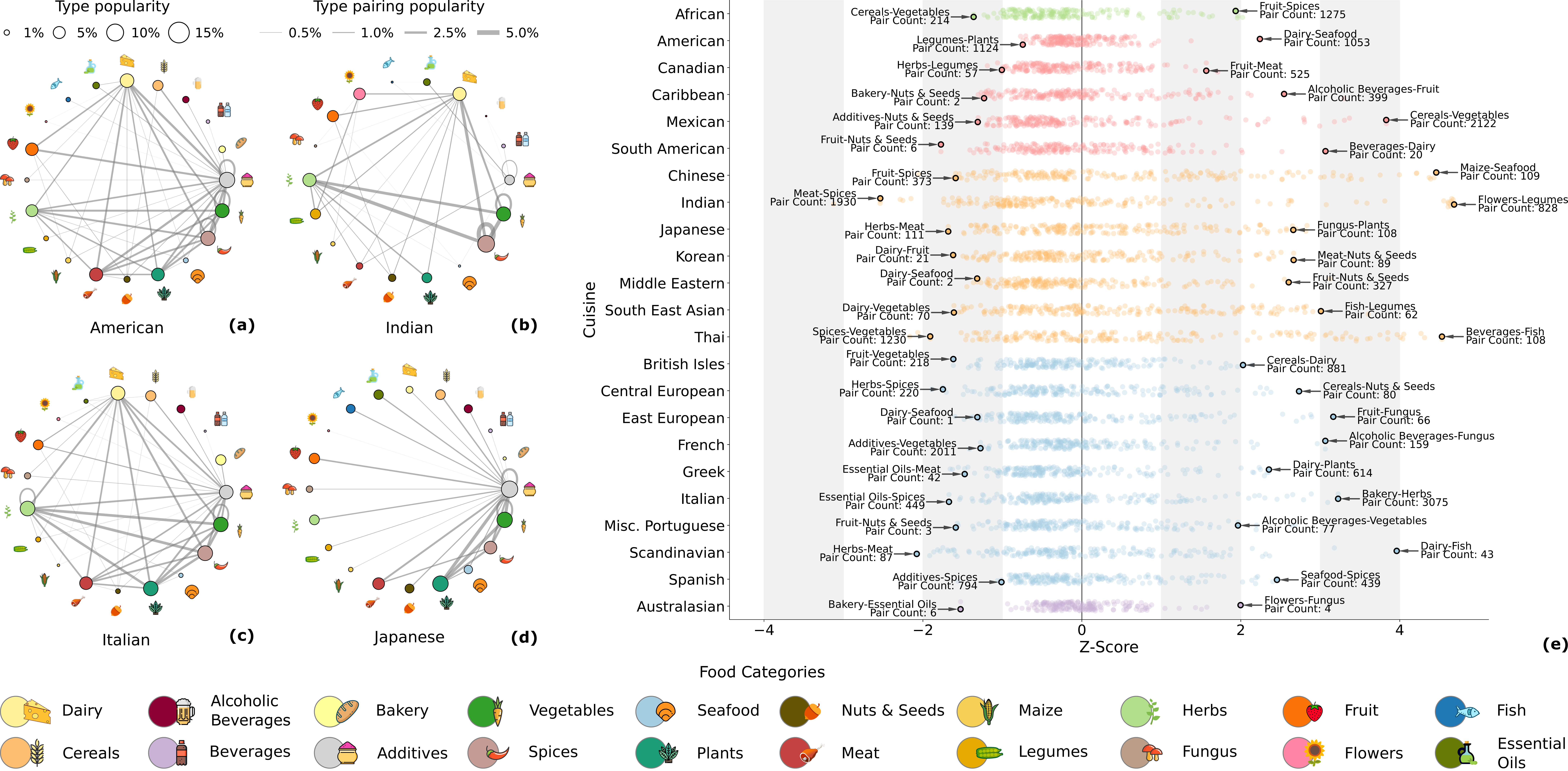}
		\caption{	\textbf{Networks of ingredient combinations and distributions of type pairing popularity across cuisines.} \textbf{(a-d)} Backbones of ingredient combinations networks for four selected cuisines: American \textbf{(a)}, Indian \textbf{(b)}, Italian \textbf{(c)} and Japanese \textbf{(d)}. Nodes correspond to ingredient types and are drawn on a circular layout with additional type label symbols, with sizes proportional to the type's recipe popularity in the cuisine. An edge is present in the network of ingredient combinations if two ingredient types occur in the same recipe at least once, and its thickness is proportional to the number of co-occurrences of types across all recipes. To highlight differences between networks we show the network backbones obtained through the disparity filter~\cite{doi:10.1073/pnas.0808904106}, plotting only statistically significant edges with a p-value larger than 0.2 \textbf{(e)} Strip plots of z-scores for edge weights across world cuisines. For each cuisine on the y-axis, the x-axis shows the z-scores of edge weights relative to the distribution of the corresponding edge weights across all 23 cuisines. A jitter is applied for better visualization. Two edges are highlighted for each cuisine, one with a high z-score and one with a low z-score, representing significant deviations in the co-occurrence patterns.}
		\label{fig:type_graph_showcase}
	\end{figure*}
	
	We constructed the ingredient type networks for each of the 23 cuisines. For better visualization of the obtained networks, due to their high density, we have used the disparity filter method introduced by Serrano et al~\cite{doi:10.1073/pnas.0808904106}. With this method, we retain only the edges whose weights represent statistically significant deviations from a null model where the weights are uniformly distributed among the links of a node. Figure~\ref{fig:type_graph_showcase}~(a-d) show the filtered ingredient networks of 4 representative cuisines obtained by setting a p-value of 0.2 (see Fig.~S3 for all other cuisines).
	We observe that the filtered networks of both American and Italian cuisines [\cref{fig:type_graph_showcase}~(a,c)] preserve a high percentage of their original edges, while also exhibiting two very similar structures. 
	Differently, Indian and Japanese filtered networks [\cref{fig:type_graph_showcase}~(b,d)] have fewer edges. 
	To gain deeper insights into how different cuisines combine food types, we analyze the $z$-scores of edge weights for specific type pairs across cuisines. For each type pair, we calculate the weight distribution across all cuisines and then derive the corresponding $z$-scores.
	In \cref{fig:type_graph_showcase}~(e), we report the scores for each cuisine using a strip plot.
	The resulting distributions of $z$-scores reveal interesting variations in the way cuisines use food type pairings. A cuisine with a long right tail in its $z$-score distribution indicates frequent use of certain type pairs compared to other cuisines. This trend is particularly noticeable in Asian cuisines, where less common ingredient pairings, such as the Flowers-Legumes combination in Indian cuisine or the Beverages-Fish pairing in Thai cuisine, are more prevalent.
	In contrast, cuisines like American, Canadian, and Australasian show a more uniform $z$-score distribution, with most scores falling within the -1 to +1 range. A likely explanation for this pattern is the history of mass immigration in these countries during the 20th century, which introduced a wide array of culinary traditions from different regions of the world. As these diverse influences blended, the distinctive characteristics of individual cuisines may have averaged out, leading to a more homogenized set of ingredient pairings.
	
	To investigate the most important associations of food pairs for each of the 23 cuisines, we report in Table~\ref{tab:top5edges} the 5 edges with the largest weight, along with their corresponding weight fraction. 
	Notably, for most cuisines, the top 5 edges (about $3\%$ of the total number of edges) represent approximately $15\%$ of the total weight.
	Additionally, the top 50 edges (about $25\%$ of the total number of edges) capture about 75\% of the total weight. The exception to this trend is observed in East Asian cuisines, where sensibly lower values are observed.
	\begin{table}
		\includegraphics[width=\textwidth]{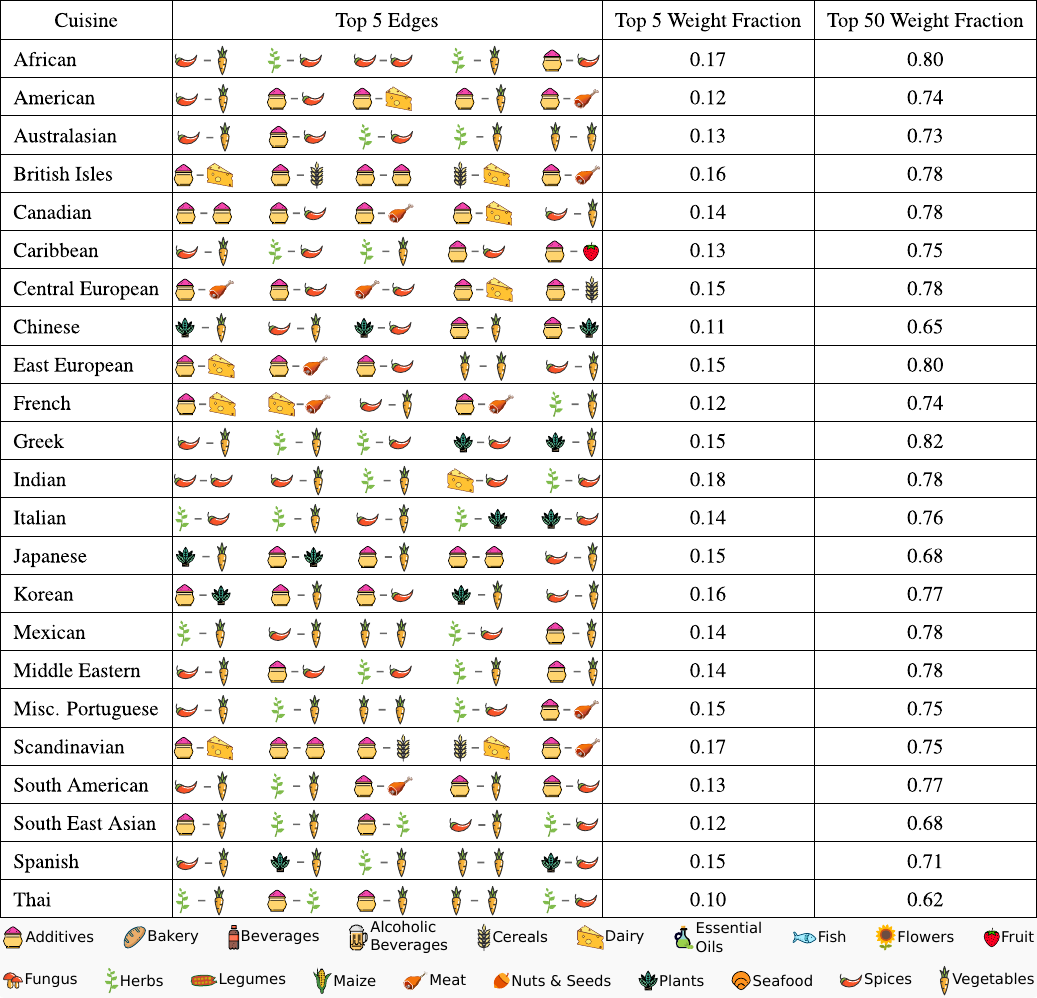}
		\caption{The second column reports the top 5 ingredient type-type pairs with the largest weight for each of the 23 cuisines. The third and the fourth columns report the total weight fraction accounted for by the top 5 and 50 type-type pairs with the largest weight, respectively.}
		\label{tab:top5edges}
	\end{table}
	\begin{table}[!htb]
		\begin{minipage}{.5\linewidth}
			\centering
			\begin{tabular}{ccc}
				\toprule
				\textbf{Type Pair} & \textbf{Average Weight} & \textbf{St. Dev.} \\
				\midrule
				Spices - Vegetables & 0.027 & 0.006 \\
				Additive - Spices & 0.025 & 0.004 \\
				Additive - Vegetables & 0.023 & 0.005 \\
				Additive - Meat & 0.022 & 0.007 \\
				Herbs - Vegetables & 0.022 & 0.008 \\
				Vegetables - Vegetables & 0.022 & 0.005 \\
				Herbs - Spices & 0.021 & 0.008 \\
				Plants - Vegetables & 0.021 & 0.006 \\
				Meat - Spices & 0.020 & 0.004 \\
				Plants - Spices & 0.020 & 0.006 \\
				\bottomrule
			\end{tabular}
		\end{minipage}%
		\begin{minipage}{.5\linewidth}
			\centering
			\begin{tabular}{ccc}
				\toprule
				\textbf{Type Pair} & \textbf{St. Dev.} & \textbf{Average Weight} \\
				\midrule
				Additive - Dairy & 0.013 & 0.020 \\
				Cereals - Dairy & 0.012 & 0.014 \\
				Dairy - Meat & 0.011 & 0.016 \\
				Additive - Cereals & 0.010 & 0.017 \\
				Dairy - Spices & 0.009 & 0.016 \\
				Additive - Additive & 0.008 & 0.020 \\
				Herbs - Spices & 0.008 & 0.021 \\
				Cereals - Meat & 0.008 & 0.013 \\
				Herbs - Vegetables & 0.008 & 0.022 \\
				Additive - Meat & 0.007 & 0.022 \\
				\bottomrule
			\end{tabular}
		\end{minipage} 
		\caption{\textbf{(Left)} Top 10 ingredient type-type pairs with the largest average weight. Corresponding standard deviations are reported in the last column. \textbf{(Right)} Top 10 ingredient type-type pairs with the largest weight variation (standard deviation). Corresponding average weights are reported in the last column.}
		\label{tab:top10_pairs}
	\end{table}
	The data in Table~\ref{tab:top5edges} reveals 
	a noticeable recurrence of certain pairs of ingredient types within the top 5 pairs of each cuisine. 
	For instance, pairings like $Spices-Vegetables$ and Additive-Vegetables frequently appear among the top 5 in many cuisines, reflecting the prominence of the three most common ingredient types (Spices, Vegetables, and Additive) across cuisines. 
	This observation is further supported by the analysis in \cref{tab:top10_pairs}~(left), where we report the 10 most frequent type pairings across all the 23 cuisines, i.e. the edges with the largest average weight. Here, Spices, Vegetables, and Additive constitute the first three edges in the ranking, underscoring their central importance across cuisines. 
	Moreover, nine of the top ten pairs, except for the pair Additive - Meat, include either Spices or Vegetables.
	The recurrence of these type-type pairs across various cuisines suggests that certain ingredient combinations are particularly versatile and essential in the construction of diverse flavor profiles. This pattern implies that, while cuisines may differ in many aspects, they share common culinary strategies that rely on a core set of ingredient pairings.
	\cref{tab:top10_pairs}~(right) reports the top 10 type pairs ranked according to the standard deviation of weights across cuisines.
	In this list, the Dairy category is involved in four of the top five pairs. This is because dairy products are essentially a staple food for the cuisines of Central and Northern Europe, while being almost completely unknown to Asian culinary traditions.
	
	\begin{figure*}[htb!]
		\centering
		\includegraphics[width=0.98\textwidth]{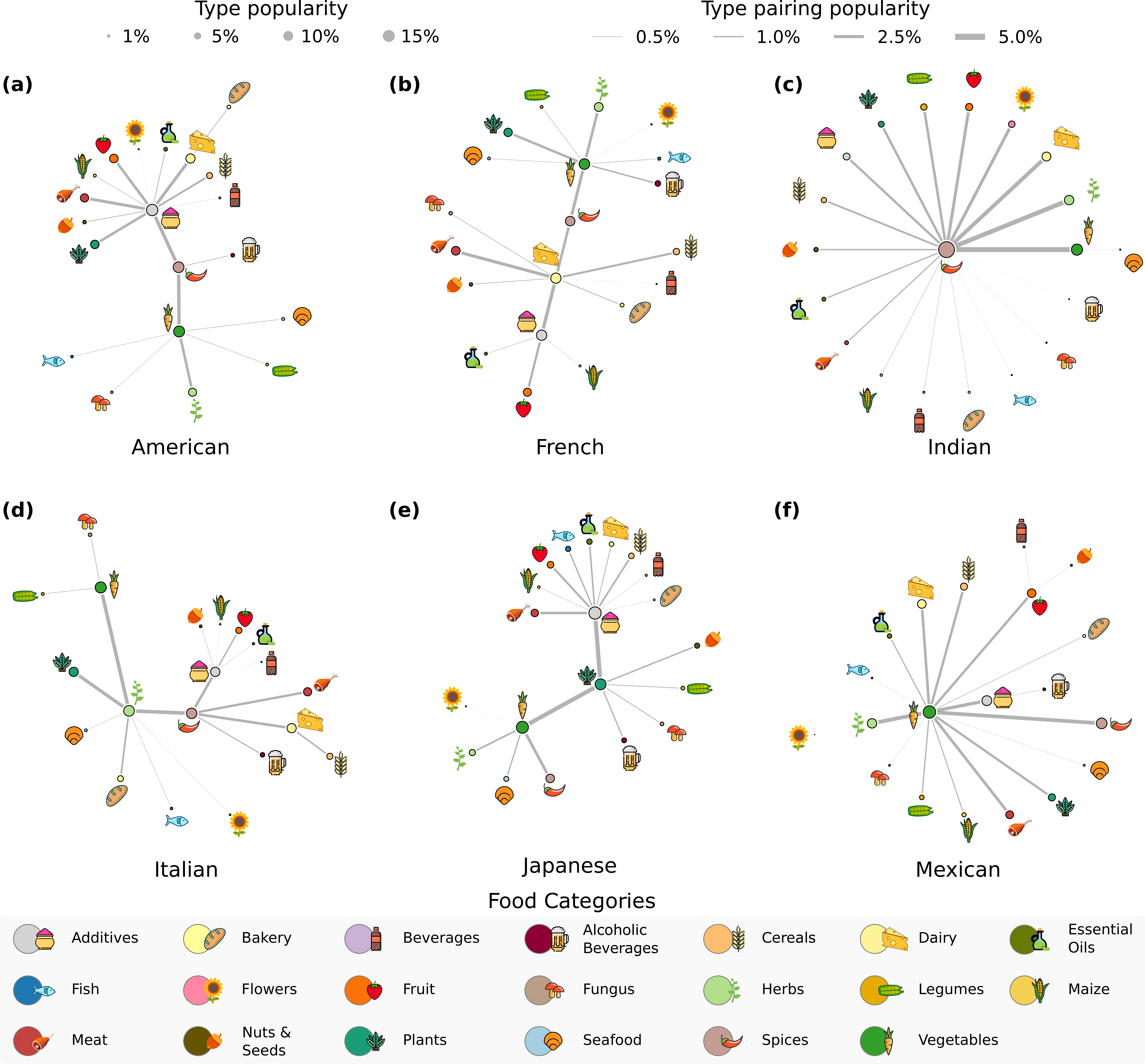}
		\caption{	\textbf{The backbone of ingredient combinations.} Maximum spanning trees (MST) of the networks of ingredient combinations for six selected world cuisines: American \textbf{(a)}, French \textbf{(b)}, Indian \textbf{(c)}, Italian \textbf{(d)}, Japanese \textbf{(e)} and Mexican \textbf{(f)}. The different topologies of the trees reveal the distinct overall organization of ingredient combinations. For instance, Indian cuisine exhibits a star-like topology centered around Spices, whereas Italian and French cuisines display a more balanced reliance on different ingredient types, such as Dairy and  Herbs. Node sizes are proportional to type popularity within the cuisine, while edge thickness corresponds to the weight in the original network of ingredient combinations, and indicates the number of co-occurrences of type pairs across all recipes.}
		\label{fig:max_span_trees_showcase}
	\end{figure*}

	To reduce the number of links 
	in \cref{fig:type_graph_showcase} and focus on meaningful associations between ingredients, we construct the  
	maximum spanning trees (MSTs)~\cite{complex_networks_latora} of the type-type graphs. 
	Indeed, MSTs are well-suited for this task as they map a graph into a connected tree. The pruned graph is obtained minimizing the number of connections while maximizing the sum of the link weights.
	MSTs can be thus regarded as connected ``minimal graphs'', easier to interpret and visualise, that give us the network fingerprint of a cuisine. The resulting MSTs for 6 of the 23 cuisines that we have studied are shown in \cref{fig:max_span_trees_showcase} (see Fig.~S4 of the Supplementary for all cuisines).
	Clear distinctions among these MSTs across cuisines underscore the effectiveness of this method in highlighting culinary differences otherwise not directly visible from the full networks. Notice that the obtained trees are all organized around one or two ``central'' nodes, i.e. the ones with the highest closeness centrality. 
	For example, the Indian MST (\cref{fig:max_span_trees_showcase}(c)) appears as a star-like graph, with Spices at its core; this configuration effectively captures the predominant centrality of Spices in Indian cuisine, which are an essential component to most of its recipes. These findings are in line with those of Jain et al.~\cite{Jain2015SpicesFT}, who demonstrated the central role of Spices through the analysis of different regional cuisines of India. The negative food pairing tendency found across Indian cuisines, that is the repeated association of complementary ingredients sharing a low number of compounds, can indeed be attributed to the ubiquitous presence of Spices. We find a similar structure also in the MST describing the Mexican cuisine (\cref{fig:max_span_trees_showcase}(e)), where Spices are replaced by Vegetables. A different picture emerges instead from some other cuisines, that feature more than one ingredient type at their core. One example is the prominent role of cheese and dairy products in French cuisine (\cref{fig:max_span_trees_showcase}(b))
	showing high affinity with certain ingredient types, like meat and mushrooms (\textit{Fungi}), traditionally associated with umami flavor. 
	The MST of the Italian cuisine (\cref{fig:max_span_trees_showcase}(d)) stands out for the central role played by Herbs (jointly with Spices) a unique feature shared only with Thai cuisine (see Fig.~S4 of the Supplementary). 
	A simple measure exemplifying structural differences across MSTs is their diameter, which quantifies how far apart the two most distant ingredient types are when moving along the shortest paths.
	In networks with small diameters, central ingredient types ---typically those with the largest frequencies--- tend to participate in a larger number of links, suggesting that these types are more evenly distributed across recipes compared to MSTs with larger diameters. Indian cuisine (\cref{fig:max_span_trees_showcase}(c)) presents the smallest diameter, corresponding for example to the 3 links connecting seafood to meat passing through Spices; among the MSTs shown here, American cuisine displays the largest diameter, corresponding for example to the 5 links connecting mushrooms to bakery via Vegetables, Spices, Additives, and Cheese. The highest diameter, formed by 6 links, is the one associated with Central European cuisine (see Fig.~S4).
	
	\subsection*{Identifying a cuisine from a set of recipes}
	
			\begin{figure*}[ht!]
		\centering
		\includegraphics[width=0.98\textwidth]{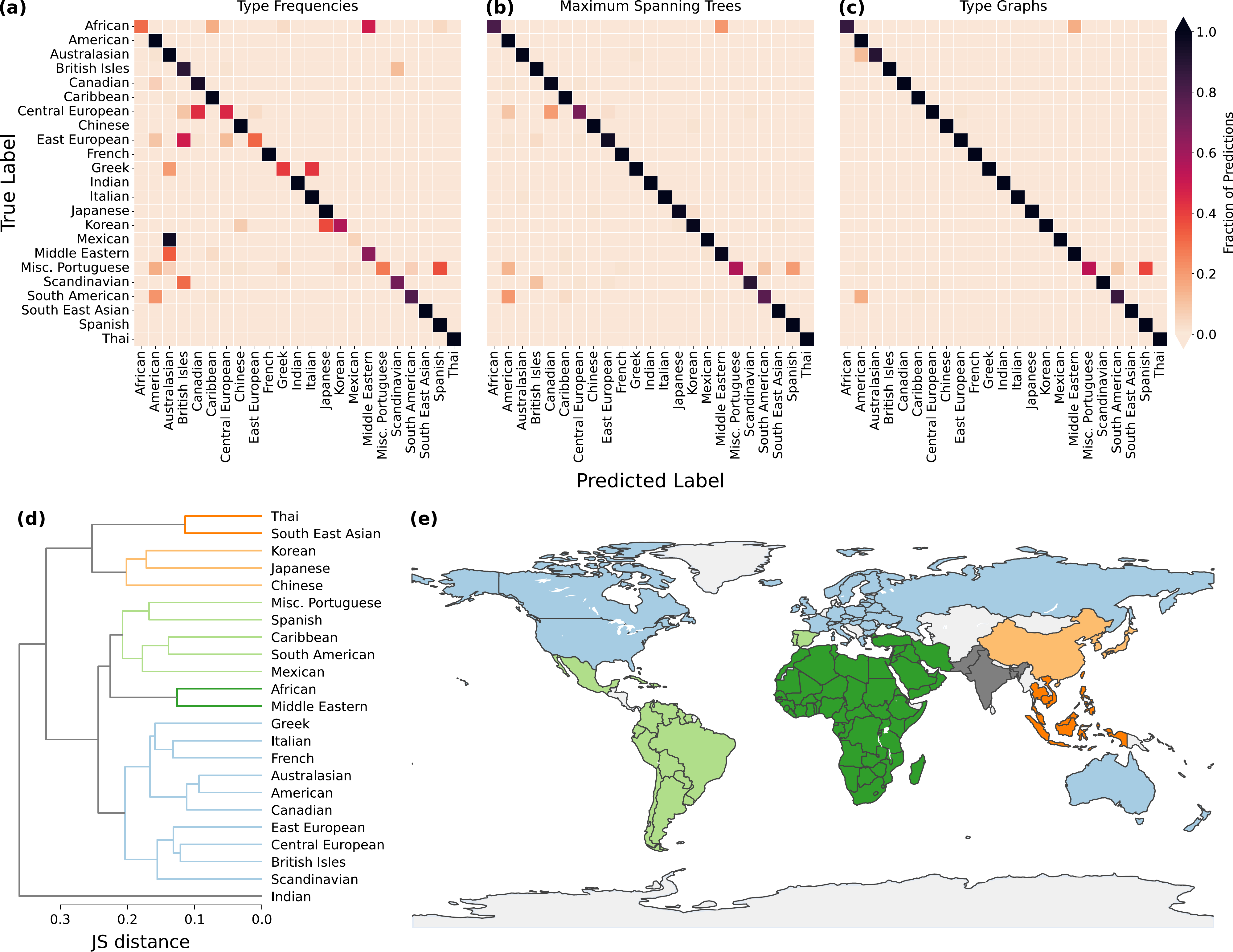}  
		\caption{	\textbf{Identification and clustering of world cuisines}. \textbf{(a-c)} Results of a Support Vector Machine (SVM) prediction model trained to associate a given set of recipes to the world cuisine they are extracted from. We report the confusion matrices obtained by using respectively: \textbf{(a)} ingredient frequency vectors, \textbf{(b)} Minimum spanning tree (MST) of ingredient combination networks, and \textbf{(c)} ingredient combination networks, as inputs to the SVM. \textbf{(d-e)} Hierarchical clustering of world cuisines based on their ingredient combination networks. \textbf{(d)} Hierarchical tree obtained from an average linkage clustering method using a Jensen-Shannon (JS) distance. Cutting the tree at distance 0.21 produces six different groups, shown with different colors. \textbf{(e)} Geographical mapping of the resulting clusters from (d) highlighting the geographical distribution of the culinary groups.}
		\label{fig:hierarchical_clustering_confusion_matrices}
	\end{figure*}
	
	We investigate here how well a world cuisine can be represented by its network of ingredient type combinations. We do this by setting up a series of classification experiments in which the task is to correctly associate a given set of recipes to the world cuisine from which they are extracted. 
	The idea is to train a classification model on three different inputs of increasing complexity, namely the vector of ingredient (type) frequencies, the MST of ingredient (type)  combinations, and the full network of ingredient (type) combinations. 
	
	In this way, it is possible to compare the predictive power of a cuisine representation based on the network of ingredient combinations, with respect to the other representations (ingredient frequencies, or MST of ingredient combinations).
	
	For each of the three different representations, we independently train a Support Vector Machine (SVM) to predict each cuisine from a sample consisting of 15\% of its recipes.  
	More details on the bootstrapping procedure and experiment pipeline can be found in the Methods section. 
	The SVMs trained on the ingredient combination networks and their corresponding MSTs achieve an average classification accuracy of 0.95 (standard deviation 0.05) and 0.87 (standard deviation 0.06), respectively. Both graph-based representations outperform the SVM trained on ingredient frequencies, which only reaches an average accuracy of 0.79 (standard deviation 0.07). To get further insights on the classification performances beyond the accuracy scores,  \cref{fig:hierarchical_clustering_confusion_matrices}(a-c) report the average confusion matrices. 
	We notice that models trained on ingredient frequencies mis-classify more cuisines than the other two models trained on ingredient combinations. 
	In particular, the heatmap in \cref{fig:hierarchical_clustering_confusion_matrices}~(a) shows that 
	cuisines belonging to either culturally or geographically close countries 
	are often mixed, e.g. African with Middle Eastern, Greek with Italian, Korean with Japanese, and Portuguese with Spanish.
	By contrast, the darkest colors along the diagonals of the heatmaps in \cref{fig:hierarchical_clustering_confusion_matrices}~(b) and (c) show that, for both MST and network-based models, the most frequently predicted cuisine label always coincides with the true label. While a degree of ambiguity persists for some cuisines when using MSTs, this is almost entirely resolved with the use of the full network of ingredient (type) combinations. 
	The only exception is the Misc. Portuguese cuisine, which is also the cuisine with the fewest number of recipes. 
	Overall, the high classification accuracy achieved by the SVM based on the ingredient combination networks demonstrates their ability to efficiently encode rich information about cuisines. 
	Their good performance suggests that networks of ingredient (type) combinations offer a more precise representation of a cuisine than the frequency of ingredients (type) alone. 
	Moreover, it is worth stressing that even though MSTs are obtained by pruning 90\% of the original edges, they are still able to achieve very high classification accuracy, further confirming the relevance of a networked approach.

	\subsection*{Classification and clustering of world cuisines}
	
	We now exploit the networks of ingredient type combinations to uncover similarities and differences among 23 worldwide cuisines and to classify them into a smaller number of groups. 
	We have adopted a hierarchical clustering based on the weighted adjacency matrices of the networks of ingredient types. We used the 'average’ linkage method and Jensen-Shannon distance to quantify how ``distant'' cuisines are from each other. 
	The full dendrogram is reported in 
	\cref{fig:hierarchical_clustering_confusion_matrices}~(d). 
	Cutting at 0.21, the ingredient type graphs identify 6 main clusters, reported with different colors. 
	Of these 6 groups, one uniquely identifies India, as proof that Indian cuisine has its unique way of combining ingredients in recipes.
	Mapping these culinary groups geographically, as illustrated in \cref{fig:hierarchical_clustering_confusion_matrices}~(e), reveals three broader geographical clusters.  Of these, one entirely coincides with Asian cuisines, further divided into two sub-clusters based on latitude (orange colors). Differently,  the remaining clusters do not exhibit clear geographical separations (light green and light blue colors).
	However, the cluster including Central and Northern Europe (light blue color) encompasses Greek, English, and Italian cuisines, as well as what are referred to as ``new world'' countries. In a way, this cluster reflects the historical and cultural influences shaped by colonialism in the evolution of world cuisines.
	This is also evident in the cluster that includes Spanish and Portuguese cuisines (light green), grouped with Latin American countries, formerly part of their colonial empires. Similarly, the presence of North Africa and the Middle East, easily explained by geographical continuity with the Iberian peninsula, could be the result of the long-lasting effects of medieval Islamic colonization of Southern Europe. Taken together, the overall clustering appears to depend on the interplay of geographical and cultural ties, which concur in shaping the ingredient profile of a cuisine.

	\section*{Discussion}
	World cuisines embody a broad and intricate fusion of flavors, traditions, and cultural influences. Over time, this elaborate composition has garnered growing attention, not only due to its implications for health and dietary practices~\cite{Cena2020-gb, Vajdi2020}, but also for its role in inspiring new fields like computational gastronomy~\cite{goel2022computational}. 
	In this work, we have introduced a novel network-based approach that sheds light on both general trends and distinctive features of culinary traditions. 
	
	We first found that the number of recipes within a single cuisine exhibits an exponential relationship with the number of ingredients. Since the existing recipes constitute only a fraction of the potential ingredient combinations 
	(as from other recipe repositories~\cite{10.1093/oxfordhb/9780199729937.001.0001, bien-etal-2020-recipenlg}, only about $10^6$ recipes exist out of an estimated $\sim 10^{15}$ potential combinations). This perceived limitation may embody a principle of least energy. It suggests that cuisines often rely on variations of successful dishes, instead of exploring completely novel combinations of ingredients.
	As highlighted by Kinouchi et al~\cite{Kinouchi_2008}, modern cuisines undergo continual evolution, with high-fitness ingredients replacing lower-fitness ones. More recently, Nazari et al.\cite{nazari2024evolution} provided direct evidence supporting this hypothesis by reconstructing the evolutionary history of Italian stuffed pasta. Factors such as antimicrobial properties have been suggested to influence ingredient selection over time~\cite{billing1998antimicrobial,sherman2001why}.
	Consequently, within a single cuisine, the space of potential ingredient combinations, even within the bounds of "gastronomical rules", remains only partially explored. This ongoing evolutionary process contributes to the dynamic and adaptable nature of culinary traditions.
	
	By examining how cuisines utilize ingredients of different types we also unveiled a pattern in how ingredient types are organized based on average prevalence. Cuisines generally tend to use the same ingredient types with comparable frequencies, suggesting a hierarchy. Categories like Vegetables, Spices, and Additive are extensively used across all cuisines, while others like Flowers, Beverages, and Maize exhibit overall low popularity. 
	We hypothesize that various factors shape this pattern. For instance, the high prevalence of types like Vegetables may stem from their nutritional profile, providing key nutrients. Similarly, the prevalence of Additive and Spices can be justified by their specific culinary functions, such as flavoring or acting as chemical agents to catalyze certain reactions or provide antimicrobial properties to food.   
	The networks of ingredient type combinations, introduced for the first time in this work, proved 
	quite effective for classification tasks, showing that world cuisines can be essentially identified by how they combine ingredients based on their types. Results presented in \cref{fig:hierarchical_clustering_confusion_matrices}
	show clearly that the unique and intertwined pattern of ingredient types constitutes a very good ``fingerprinting'' for each cuisine.
	This is also confirmed when considering dimensionality reduction methods, such as tSNE~\cite{JMLR:v9:vandermaaten08a}(see Fig.~S5 of the Supplementary for details), where ingredient type graphs form a culinary continuum that goes from East Asia to Scandinavia, where geographical proximity appears to be particularly relevant in determining the 2D projections. These findings align with those of Zhu et al.~\cite{10.1371/journal.pone.0079161}, who, employing a distinct approach based on ingredient frequencies, analyzed regional cuisines in China and identified geographical proximity as the key factor of regional cuisine similarities. 
	Furthermore, network measures derived from ingredient-type graphs (e.g., the spectral radius) show positive correlations with exogenous World Development Indicators, such as food imports and food insecurity, for the countries associated with that cuisine (see Fig.~S6). One possible explanation is that cuisines with higher spectral radii might reflect a less varied combination of ingredient types, which could correspond to more limited diets and thus correlate with higher food insecurity. Likewise, this measure may hint at shortages of certain ingredient types, potentially requiring greater reliance on imports. These preliminary observations highlight the need for further investigation to confirm and refine these relationships.
	
	Finally, we showed how the analysis of the Maximum Spanning Trees (MSTs) of the graphs of ingredient type combinations, widely used to filter networks in finance~\cite{Mantegna_1999, VALLE2018146} and biology~\cite{10.1093/bioinformatics/btp109, wu2008efficient}, can highlight the most relevant ingredient types and their pairings. 
	Even though MSTs contain less information than the original type-type graphs, they provide a compact representation in terms of the essential components defining the culinary identity of a cuisine. In this respect, MSTs not only enable us to refine the visual representation of food type graphs by minimizing complexity and redundancy, but allow for a more intuitive exploration and analysis of the relationships between ingredients.\newline
	
	Since the structure of these graphs ultimately depends on how ingredients are categorized, it is important to stress the implications and limitations of our classification choices. We acknowledge that the ingredient classification we adopted in this work inherits intrinsic arbitrariness, carried over from the original dataset~\cite{CulinaryDB}. First, the choice of categories may reflect potential cultural biases: ingredients that in our framework are grouped under the same broad categories might be perceived as distinct food types in certain cultures. A notable example is that of seaweeds and algae, which are categorized under Seafood in our dataset based on their environmental (marine) origin. While this classification is consistent with ecological criteria and taxonomic ambiguity (algae are not strictly plants), it may overlook their cultural specificity, such as the central role of seaweeds in Japanese cuisine, potentially motivating their treatment as a separate ingredient category. This example highlights a broader point: culturally specific ingredient types offer a compelling reason to explore alternative classification systems, especially in regionally focused or cuisine-specific studies, where finer distinctions may lead to more meaningful insights. A second simplification in our framework concerns food processing: we merge different processed forms of the same raw item (e.g., fresh, dried, or canned tomatoes) into a single entity. Consequently, our categories capture the natural origin of ingredients but not the form they take after processing. While this choice allowed a first analysis by reducing complexity and assumptions, we acknowledge that processing can significantly influence ingredients' function and flavor, particularly in certain cuisines. A natural extension our our work would thus account for processing stages through a hierarchical categorization scheme, with subcategories (e.g., dried, fermented) nested within broader types (e.g., Vegetables, Dairy). Such more granular taxonomy could enrich future analyses, if based on a clear and principled annotation of the database. Moreover, our analysis relies on data from a single source. Expanding the dataset to include a broader range of sources and cuisines would improve the robustness and generality of our findings. Additionally, our analysis considered a static snapshot of cuisines, using recipes available at a single point in time. However, against the well-established idea of national cuisines as part of the immutable traditions, culinary practices continuously evolve due to various factors such as technological advancements, ingredient availability, and local trends, all pushing for innovations in different directions. Hence, there is a need to recognise cuisines as dynamic and growing collections of recipes. Implementing a temporal network analysis would help to capture the evolution of culinary trends and ingredient popularity over time, and also to test the robustness of our network-based framework. Finally, any network approach primarily focuses on pairwise interactions, namely on the combination of pairs of ingredients. However, networks may not fully capture the complexity of culinary traditions. Extending our approach to higher-order networks, such as hypergraphs, to better describe higher-order interactions~\cite{BATTISTON20201} among multiple ingredients, could provide a more nuanced understanding of ingredient combinations and their roles within recipes. 
	
	\section*{Methods}
	
	\subsection*{Dataset} 
	CulinaryDB is a large repository consisting of 45772 recipes and two associated ingredient lists: one containing 929 individual ingredients, the other 103 compound ingredients, accounting for a total of 1032 unique ingredients. Of these, only 695 appear in the recipes themselves. The data were collected by CoSyLab~\cite{culinarydb_webpage} from the following platforms: AllRecipes~\cite{allrecipes}(16177), Food Network~\cite{foodnetwork}(15917), Epicurious~\cite{epicurious}(11069) and TarlaDalal~\cite{tarladalal}(2609). Recipes are grouped into 26 cuisines represented by world regions arching over 5 continents. A world region can either coincide with the national borders of a state or represent a broader geocultural area, such as Central America and Scandinavia (see Supplementary Note~1 and Table~S2 for a detailed description of world regions). In the data set, each recipe is structured as a set of ingredients, which are classified into $T=20$ different culinary \textit{categories}, or \textit{food types}, namely: Additive, 
	Bakery, 
	Beverages,
	Alcoholic Beverages,
	Cereals, 
	Dairy,
	Essential Oils, 
	Fish,
	Flowers,
	Fruit,
	Fungi,
	Herbs,
	Legumes,
	Maize,
	Meat,
	Nuts \& Seeds,
	Plants,
	Seafood,
	Spices,
	Vegetables. 
	Although this classification was defined by the creators of CulinaryDB\cite{CulinaryDB}, a comprehensive inspection shows that it follows a coherent set of principles grounded in biological origin, culinary function, and degree of processing. Ingredients are primarily categorized by their natural source, such as Meat (terrestrial animals), Fish (aquatic vertebrates), Seafood (aquatic invertebrates), Plants (leaves, stems, and roots), or Fungi, and secondarily, when applicable, by their dominant culinary role. For instance, aromatic leaves are labeled as Herbs, flavoring seeds as Spices, and concentrated extracts as Essential Oils. Items resulting from substantial transformation (e.g., Dairy, Bakery, Beverages) or used predominantly as process aids (e.g., Additives) are assigned to distinct classes. Borderline cases are resolved based on prevailing culinary identity: for example, honey and soy sauce, though processed, are categorized as Plants rather than Additives due to their natural origin and frequent culinary use. Manual inspection of all standardized ingredients confirms consistent application of these criteria.

	For our analysis, we used data from 23 of the 26 cuisines, discarding those with fewer than 100 recipes to ensure statistical reliability, ending with a total of $R=45661$ recipes. Additionally, we performed a preprocessing step to standardize and simplify the set of 695 ingredients found in recipes, resulting in a reduced set of $I = 604$ distinct ingredients. This included merging different shapes or varieties (e.g., all pasta types as ‘pasta’), unifying fresh, dried, or processed forms (e.g., dried and fresh peas as ‘peas’), and generalizing ingredient subtypes (e.g., cherry, plum, and globe tomatoes as ‘tomatoes’). We also aggregated ingredients produced through similar transformative processes into broader categories (e.g., all cheeses as ‘cheese’, all grape-based fermented drinks as ‘wine’), and grouped different animal parts under their corresponding species (e.g., chicken wings, thighs, and liver as ‘chicken’). A detailed description of the preprocessing rules and illustrative examples is provided in Supplementary Note~1.

	For each cuisine $c$, with $c=1,\ldots,23$, we denote as ${\cal R}_{c}$ the set of recipes and, as $R_c = | {\cal R}_{c}|$, the number of recipes in cuisine $c$, representing the size of the cuisine. 
	Each recipe $r \in {\cal R}_{c}$ can be viewed as a subset of the finite set of   ingredients ${\cal I}_{c}$ used in cuisine $c$, from which dishes are created through different combinations. 
	We indicate the number of ingredients in recipe $r$ as $n_r$, and the number of ingredients in cuisine $c$ as $I_c = | {\cal I}_{c} |$, with $I_c \le I$. We indicate an ingredient type as $t$, with $t = 1, \ldots, T$, and as $t_{i}$ the type of an ingredient $i$, with $i=1,2,\ldots, I_c$. 
	We define the popularity $f_c(t)$ of an ingredient type $t$ in cuisine $c$ as the fraction of ingredients of type $t$ in the recipes of cuisine $c$. To calculate $f_c(t)$, we sum the number of ingredients of type $t$ across all the recipes in cuisine $c$, and normalize this sum by the total number of ingredients in all recipes of cuisine $c$.  
	Similarly, we define the popularity $p_c(t)$ of an ingredient type $t$ across the recipes of cuisine $c$, as the fraction of recipes of $c$ containing at least an ingredient of type $t$.
	In other words, $f_c(t)$ represents the probability that a randomly extracted ingredient from recipes of cuisine $c$ belongs to type $t$. Differently, $p_c(t)$ is the probability that a randomly extracted recipe of cuisine $c$ contains at least an ingredient of type $t$. 
	Formally, these definitions can be expressed as follows:
	\begin{equation}
		f_c(t) = \frac{\sum_{r \in \mathcal{R}_{c}} \sum_{i \in r} \delta(t_i, t)}{\sum_{r \in \mathcal{R}_{c}} n_r},
	\end{equation}
	and 
	\begin{equation}
		p_c(t) = \frac{\sum_{r \in \mathcal{R}_{c}} \mathbf{1}\left(\exists i \in r \mid t_i = t\right)}{R_c} 
	\end{equation}
	where $\delta (a,b)$ is the Kronecker delta function, which is 1 if $a=b$ and 0 otherwise, and $\mathbf{1}(\bullet)$  is the indicator function that is 1 if any ingredient in the recipe belongs to type t and 0 otherwise.
	
	\subsection*{Network representations of the data}
	
	We represent the data using three different network structures. The first is the \emph{Recipe-Ingredient graph}, which is a bipartite graph~\cite{newman2006structure} representation of the recipes of a cuisine $c$.  The node sets of the graph consist of recipes and ingredients appearing in cuisine $c$. Specifically, if an ingredient $i$ is part of a recipe $r$, an undirected edge connects $i$ to $r$. The obtained graph is thus bipartite, as there are no edges inside the two separate sets of ingredients and recipes. It is straightforward to show that the degree distribution $P_r (k)$ for the recipes coincides with the recipe size distribution $P_r (n)$.
	
	The second representation is the Ingredient-Ingredient graph, defined as the weighted projection~\cite{doi:10.1073/pnas.0400087101} of the recipe-ingredient graph on the set of ingredients ${\cal I}_{c}$. A pair of ingredient nodes $i$ and $j$ is connected if there exists at least one recipe containing both $i$ and $j$. The weight $w_{ij}$ is equal to the number of recipes of the cuisine that uses both $i$ and $j$ as ingredients.
	
	The third representation is based on ingredient type graphs. To construct the type graph for a cuisine $c$ we employ the following procedure. For each recipe $r$ within the cuisine, we construct the multi-set~\cite{hein2003discrete} of type pairs $t_i, t_j$ for every unordered pair of ingredients $i, j$. We then compute the co-occurrence frequency $n(t_i, t_j)$ of each type pair $t_i, t_j$, across the multi-sets. We add the undirected edge $(t_i, t_j)$ to the network, with the edge weight $w_{t_i, t_j}$ set equal to the computed co-occurrence frequency.
	We finally enrich the network nodes with relative abundance ($v_t$) of the ingredient types, computed as the fraction of recipes that contain at least one ingredient of type $t$. 
	Figure~\ref{fig:type_graph_showcase}~(a-d) shows the backbones of the resulting food type networks for 4 of the 23 worldwide cuisines. Node sizes in the plots represent relative frequencies ($v_t$) of different food types, with larger nodes indicating more prevalent food categories within a cuisine. \\
	See Supplementary Note~1 and Fig.~S7 for a graphical representation. 
	
	\subsection*{Classification experiment}
	The classification experiment comprises 250 repeated classification tasks, each involving a different data set comprising 23 classes corresponding to the different cuisines. Each data point in our dataset is represented as a pair $(X, y)$, where $X$ denotes one of the three possible feature representations of a recipe (type frequency vectors, ingredient type graphs, and their respective MSTs), and $y$ is the corresponding cuisine label.

	For each cuisine, we reserve 85\% of the available recipes for training and leave the remaining 15\% for testing. To create the training set, we follow a systematic sampling approach to ensure a balanced representation across all classes. For each of the 23 cuisines, we sample 75 recipes from the 85\% reserved for training and use these to generate the feature representation $X$, obtaining a data point $(X, y)$ for each of the three feature representations, with $y$ being the associated class (cuisine) label. We repeat this process 100 times for each class (cuisine), resulting in three separate balanced training sets, each containing 2,300 data points (100 samples $\times$ 23 classes). 
	For the test set, we use the remaining 15\% of recipes to generate a single data point $(X, y)$ for each cuisine, resulting in a minimal balanced test set of size 23.

	We utilize a Support Vector Machine (SVM) classifier for the classification tasks. For each of the three feature representations, a distinct SVM model is trained on the corresponding balanced training set. The performance of each SVM model is then evaluated on the respective test set, which contains one data point for each of the 23 cuisines.

	To ensure the robustness and consistency of our results, the entire process—data preparation, training set construction, model training, and evaluation is repeated 250 times. This approach allows us to account for variability in data sampling and provides a comprehensive assessment of the classifier's performance across different feature representations and cuisine classes.

	The performance associated with each feature representation is quantified in terms of the average test accuracy of the model trained using that representation across all 250 iterations. We use the standard deviation of the test accuracy as a measure of the error. This approach allows us to evaluate the stability and reliability of the classifier's performance across multiple iterations and feature sets.
	
	\section*{Data availability}
	The data used in this manuscript are available from the original source at~\cite{culinarydb_webpage}.
	
	\section*{Code availability}
	All scripts supporting the findings of this study can be found at \href{https://github.com/claudiocaprioli/ingredient-combination}{https://github.com/claudiocaprioli/ingredient-combination}.
	
	\section*{Acknowledgements}
	A.S. acknowledges support from the SNSF COST project (grant no. IZCOZ0\_198144).
	
	\section*{Author contributions statement}
	All authors designed the study. C.C. and S.K. analysed the data. All authors interpreted the results, wrote and approved the manuscript.
	
	\section*{Competing interests} The authors declare that they have no competing interests.
	
	\section*{Correspondence} Correspondence and requests for materials should be addressed to \href{iacopo.iacopini@nulondon.ac.uk}{iacopo.iacopini@nulondon.ac.uk}, \href{andrea.santoro@centai.eu}{andrea.santoro@centai.eu}.
	
	\bibliography{sample}

	
	\pagebreak[2]
	\clearpage
	\newpage
	\startsupplement
	\renewcommand{\theequation}{Eq. S\arabic{equation}}
	\let\oldsection\section
	\renewcommand{\section}[1]{\oldsection{Supplementary Note~\thesection \ -- \ #1}} 
	
	\newpage
	
	\section{Data Preprocessing and Representation}
	
	Our study utilizes CulinaryDB, a repository of recipes introduced in the main text, encompassing 23 distinct cuisines, each representing a geocultural area. Table~\ref{tab:cusine_country} delineates the correspondence between cuisines and their respective geographical regions, providing essential context for comprehending the cultural origins and culinary traditions associated with each cuisine.
	
	To prepare the raw culinary dataset for analysis, we performed a preprocessing step to standardize and simplify the ingredients set. Our preprocessing consisted in mapping groups of similar ingredients into single ingredients, thereby reducing the overall number of ingredients. We applied the preprocessing to the following cases:
	
	\begin{itemize}
		\item \textbf{Merging Redundant Entries}: Duplicate or near-duplicate ingredient entries referring to the same entity were merged into a single standardized form. This was essential to reduce noise introduced by inconsistent naming conventions across recipes.
		\item \textbf{Combining Ingredients of Similar Shape, Color, and Taste}: Different shapes of pasta and noodles were grouped together into the categories 'pasta' and 'noodles'.
		\item \textbf{Merging Fresh and Dry Forms}: Ingredients available in fresh, dry, and refrigerated forms were combined. For example, fresh peas, dry peas, and refrigerated peas were all grouped under 'peas'.
		\item \textbf{Generalizing Ingredient Subtypes}: Specific cultivars or commercial variations were merged into their general class. For instance, cherry tomatoes, plum tomatoes, and globe tomatoes were all grouped under 'tomatoes'.
		\item \textbf{Grouping by Transformative Process}: Ingredients that result from a crafting process (e.g., fermentation, baking, curing) were grouped into broader culinary categories. For example:
			\begin{itemize}
				\item All types of cheese (e.g., mozzarella, cheddar, blue cheese) were grouped under ‘cheese’.
				\item All bread types made from cereal sources were grouped under ‘bread’.
				\item All fermented grape-based beverages (e.g., red wine, white wine, port wine) were merged as ‘wine’.
				\item All types of vinegar (e.g. balsamic, apple, rice, cider) were combined into 'vinegar'.
		\end{itemize}
		\item \textbf{Grouping Animal Parts}: Ingredients from various parts of animals were all grouped under their respective animal categories. For example, chicken wings, chicken offal, chicken livers, chicken thighs and chicken breast were all combined into 'chicken'. 
		\item \textbf{Other Examples}: All types of mushroom were merged into a single ‘mushroom’ entry.
	\end{itemize}
	
	This preprocessing was crucial to reduce the complexity of the dataset and to ensure a consistent representation of ingredients across recipes.\\
	
	To provide additional context on recipe complexity across cuisines, we report for each cuisine the recipe in our dataset that contains the largest number of unique ingredients. Table~\ref{tab:largest_recipes} lists the recipe titles, the total number of ingredients, and the full ingredient lists. These recipes illustrate the variability in ingredient diversity and reflect characteristic culinary practices within each cuisine.\\
	
\begin{figure*}[bh!]
		\centering
		\includegraphics[width=\textwidth]{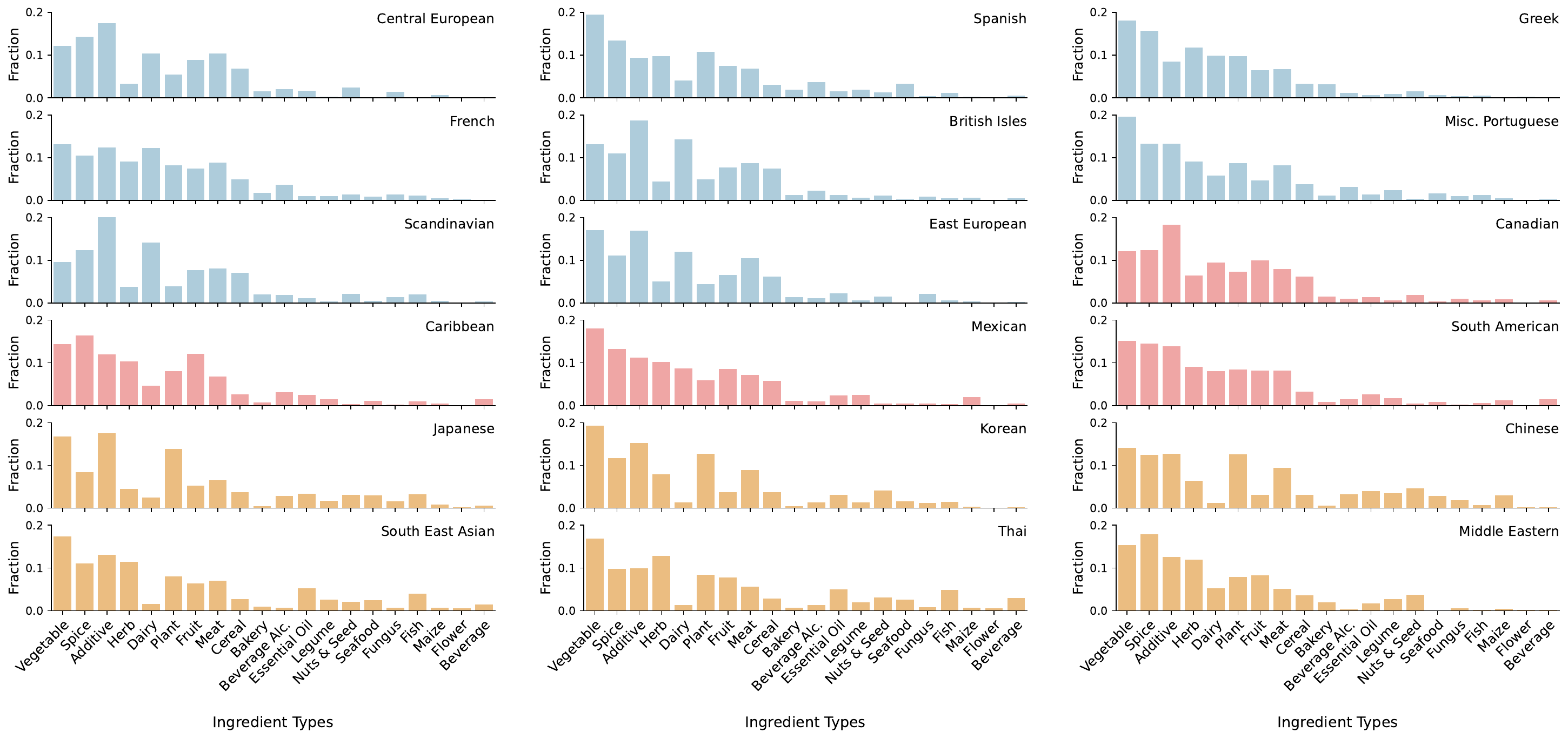}
		\caption[Ingredient Type Popularity Profile]{\textbf{Ingredient Type Popularity Profile} Bar chart representation of ingredient type popularity profiles for rest of the cuisines, highlighting the variations in ingredient preferences.}
		\label{fig:SI_ingredient_profile}
	\end{figure*}

\begin{figure*}[bh!]
		\centering
		\includegraphics[width=\textwidth]{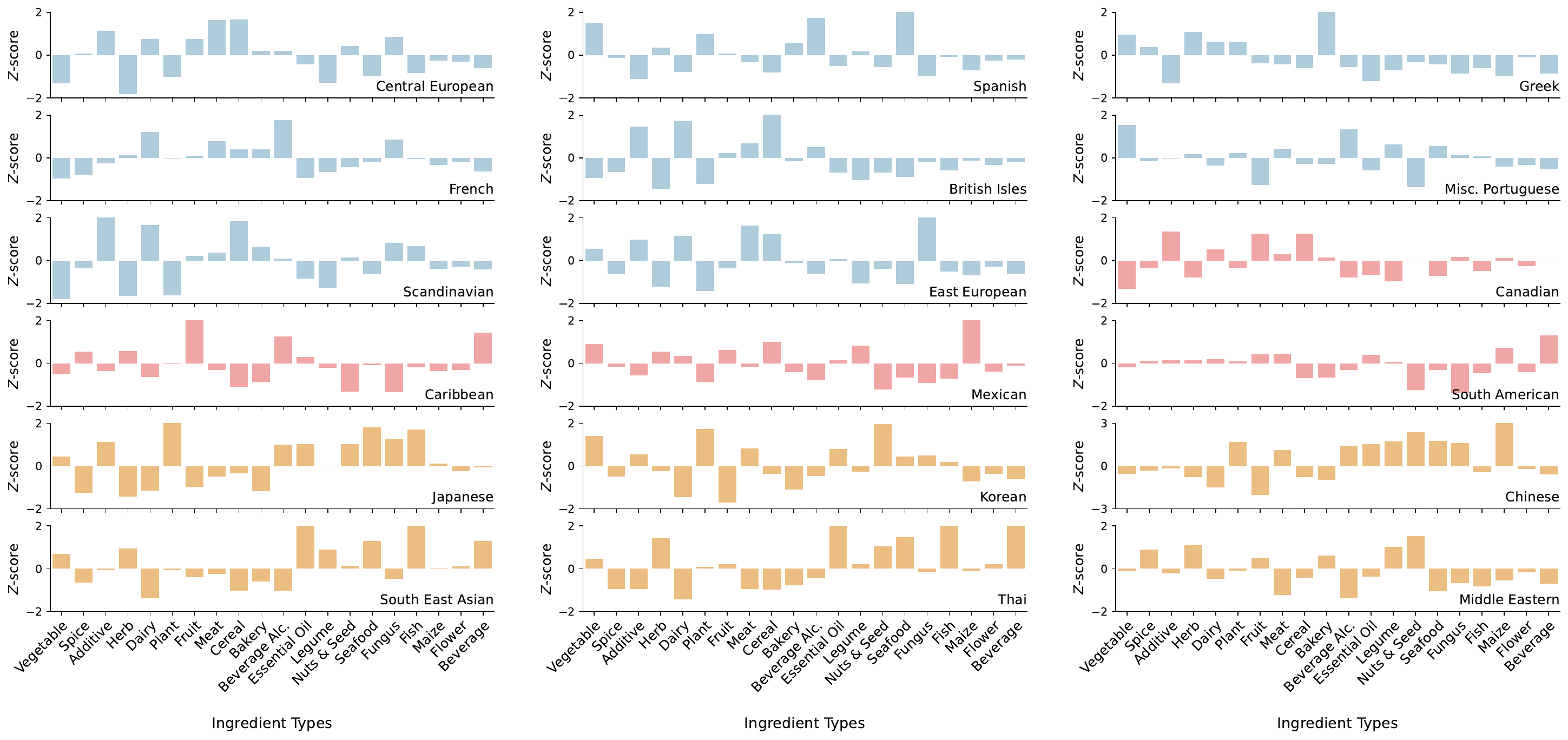}
		\caption[$Z$-score Profile]{\textbf{$Z$-score Profile} Bar chart representation of $z$-scores for rest of the cuisines, revealing the relative significance of ingredient types.}
		\label{fig:SI_z_score_profile}
	\end{figure*}

	In Fig.~2 (b) and (d) of the main text, we presented heatmap representations of both the ingredient type popularity profile and the corresponding z-scores for all 23 cuisines. Additionally, for five of these cuisines, we depicted their popularity profiles and z-scores as bar charts (Fig.~2 (c) and (e)). Here, in Figs.~\ref{fig:SI_ingredient_profile} and \ref{fig:SI_z_score_profile}, we extend this analysis to the remaining 18 cuisines.
	
	\section{Food Type Graphs}
	
	In the main text, we have introduced food type graphs. By looking at the way ingredient types are combined, such graphs represent cuisines at a coarser level of organization. Figure~\ref{fig:SI_network_representations} illustrates the three different graph representations of a cuisine: (i) bipartite recipe-ingredient network, (ii) ingredient projection network, (iii) ingredient type-ingredient type graphs. From (i) to (iii), each of these graphs correspond to a cuisine representation at an increasingly higher level of granularity. 
	
\begin{figure*}[bh!]
		\centering
		\includegraphics[width=0.985\textwidth]{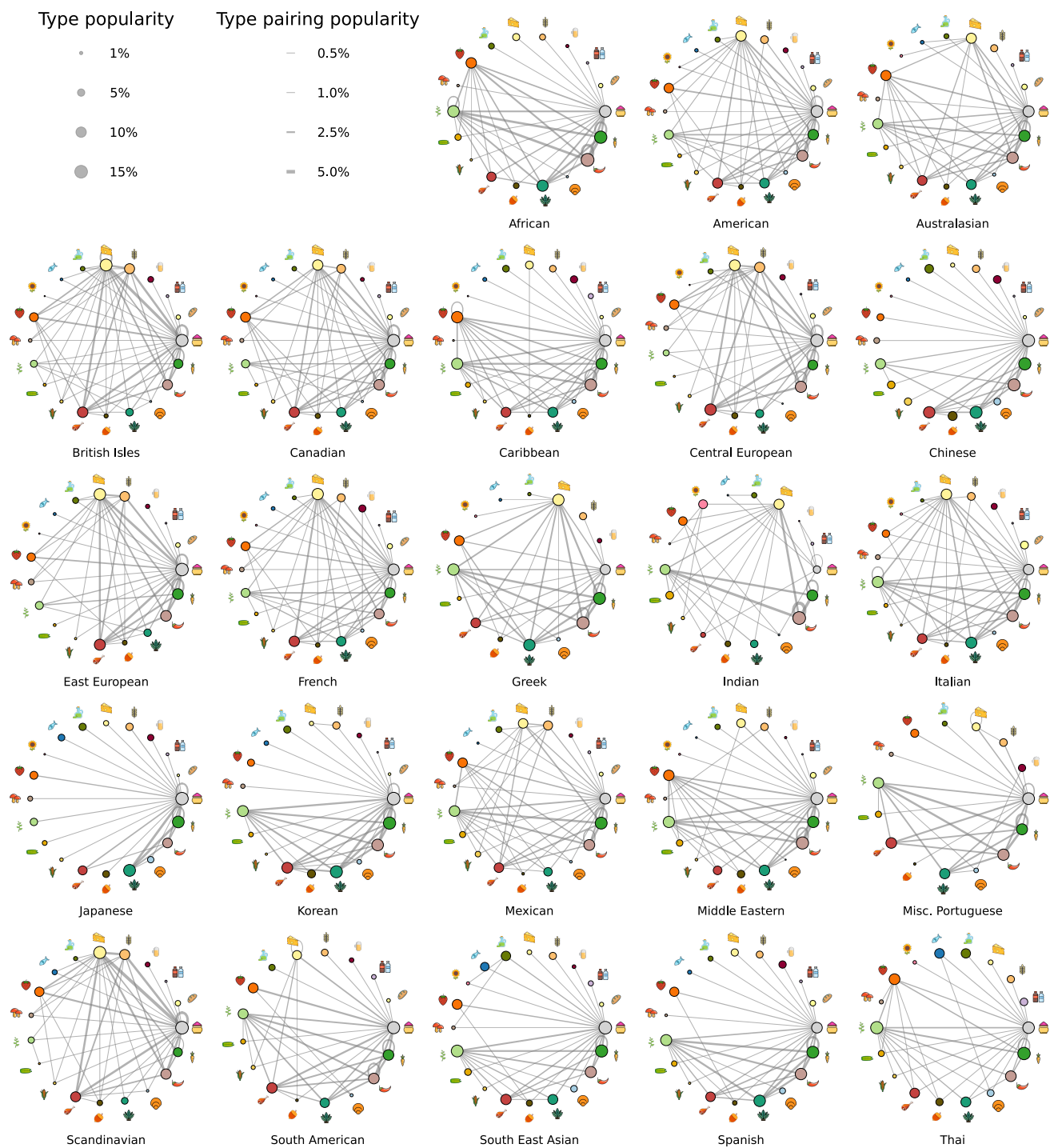}
		\caption[Networks of ingredient combinations]{\textbf{Networks of ingredient combinations.} Ingredient combination networks for all 23 analyzed cuisines, filtered using the disparity filter method introduced by Serrano et al.~\cite{doi:10.1073/pnas.0808904106}. 
			Refer to Fig.~\ref{fig:type_graph_showcase}~(a-d) for details on the realization.}
		\label{fig:SI_type_graphs_showcase}
	\end{figure*}

\begin{figure*}[bh!]
		\centering
		\includegraphics[width=0.985\textwidth]{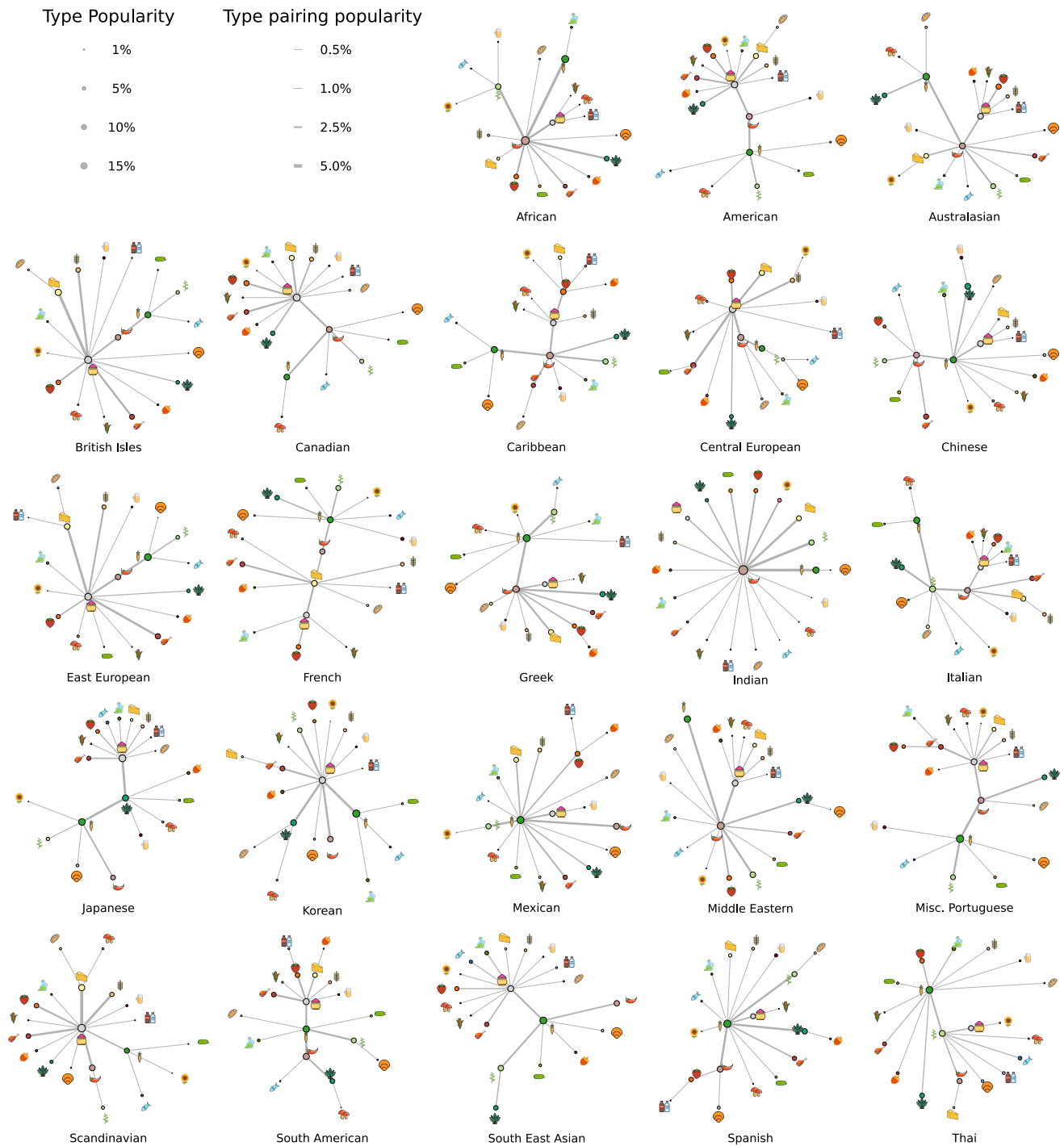}
		\caption[Maximum Spanning Trees of Ingredient Combination Networks]{\textbf{Maximum Spanning Trees of Ingredient Combination Networks}. Maximum Spanning Trees of the Ingredient Type Graphs for all 23 cuisines. Details on the realization are reported in Fig.~\ref{fig:max_span_trees_showcase} of the main text.}
		\label{fig:SI_mst_showcase}
	\end{figure*}

	Focusing on the coarsest granularity, we initially presented eight filtered food type graphs by retaining the 50 largest weighted edges to enhance visualization. Subsequently, we showcased the Maximum Spanning Trees (MSTs) for six food type graphs, providing a more distilled representation of food types connections. Here, we report the food type graphs for all the 23 cuisines. In Fig.~\ref{fig:SI_type_graphs_showcase} we show the backbones of the food type graphs, while in Fig.~\ref{fig:SI_mst_showcase} we show the Maximum Spanning Trees of the unfiltered graphs.\\
	
	Our examination of the food type graphs indicates varying levels of organization, ranging from single polarity to multi-polarity. Some graphs exhibit single polarity with a centralized structure, where a specific food category, such as in Indian and Mexican cuisines, serves as a central hub with numerous connections. In contrast, others demonstrate multi-polarity, featuring a more balanced structure with multiple central nodes. Examples of this balanced organization include French, Italian, and Japanese cuisines, where diverse food categories collectively contribute to the overall culinary identity. 
	
	Notably, we observe that the structure of each Maximum Spanning Tree is highly variable, suggesting that MSTs have the ability to highlight distinctive features of individual cuisines that may not be readily apparent from unfiltered food type graphs. 
	However, we are still able to observe common patterns across the Maximum Spanning Trees. First, the Spices category emerges as a central node in the trees of numerous cuisines, hinting at its universal importance in defining culinary profiles. Additionally, Northern European cuisines consistently feature the Additive food type as the most central node across their respective Maximum Spanning Trees, suggesting a shared emphasis on this category within this culinary group. \\
	
\begin{figure*}[bh!]
		\centering
		\includegraphics[width=0.985\textwidth]{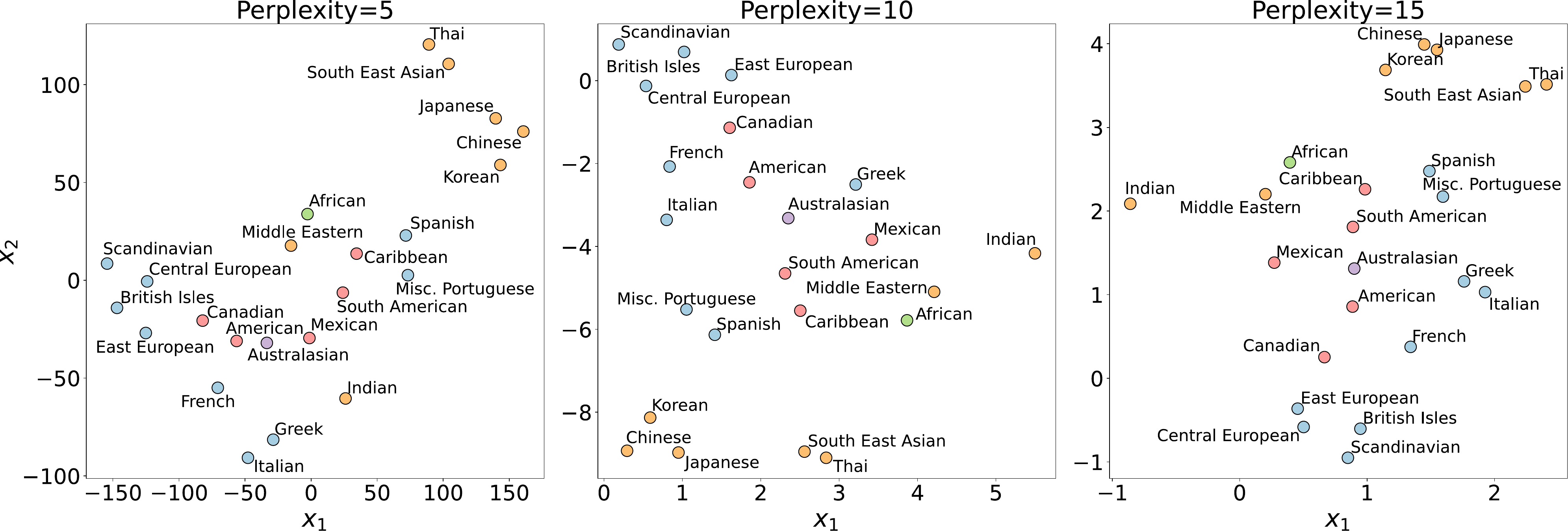}
		\caption[2d t-SNE plots of ingredient type graphs]{\textbf{2d t-SNE plots of ingredient type graphs}. Each plot represents the 2 dimensional embedding of the weighted adjacency matrices of type graphs for all 23 cuisines. (a) Perplexity 5, (b) Perplexity 10, (c) Perplexity 15. Cuisines are colored according to their continent affiliation for additional context. Asian and Northern European cuisines exhibit clustering across all perplexity levels, indicating significant similarities in their ingredient usage.}
		\label{fig:SI_tSNE}
	\end{figure*}

	To provide a geometrical picture of the proximity of food type graphs, we perform a 2D t-SNE analysis on the weighted adjacency matrices. Figure~\ref{fig:SI_tSNE} illustrates the results for three different perplexity values. The projections appear compatible with the results of the hierarchical clustering performed in the main text, offering a consistent geometric representation of culinary relationships.
	
	The 2D t-SNE unveils a culinary continuum, positioning Northern European and East Asian cuisines at opposite extremes, at all perplexity levels. This continuum reflects the inherent similarities and differences in the culinary characteristics of these regions, emphasizing the utility of t-SNE in capturing the geometric relationships between food type graphs.\\
	
	To further demonstrate the efficacy of ingredient type graphs in capturing information about the geocultural regions they represent, we selected two food-related World Bank development indicators: Food Imports and Moderate to Severe Food Insecurity. Food Imports is defined as the percentage of merchandise imports comprising food items. Specifically, it includes commodities classified under the Standard International Trade Classification (SITC) sections 0 (food and live animals), 1 (beverages and tobacco), 4 (animal and vegetable oils and fats), and division 22 (oil seeds, oil nuts, and oil kernels). Moderate to Severe Food Insecurity refers to the percentage of the population experiencing moderate or severe limitations in accessing adequate food. This indicator is calculated as a three-year average to account for annual fluctuations. These indicators are defined at the country level. To aggregate them for our defined regions of \ref{tab:cusine_country}, we employed weighted calculations: for Food Imports, we computed a weighted sum where the weights correspond to each country's merchandise imports. This approach ensures that countries with higher import volumes have a proportionate influence on the regional indicator. For Moderate to Severe Food Insecurity, we calculated a weighted sum using each countries populations as weights, thus accounting for the varying population sizes and providing a more accurate regional representation.
	
\begin{figure*}[bh!]
		\centering
		\includegraphics[width=0.985\textwidth]{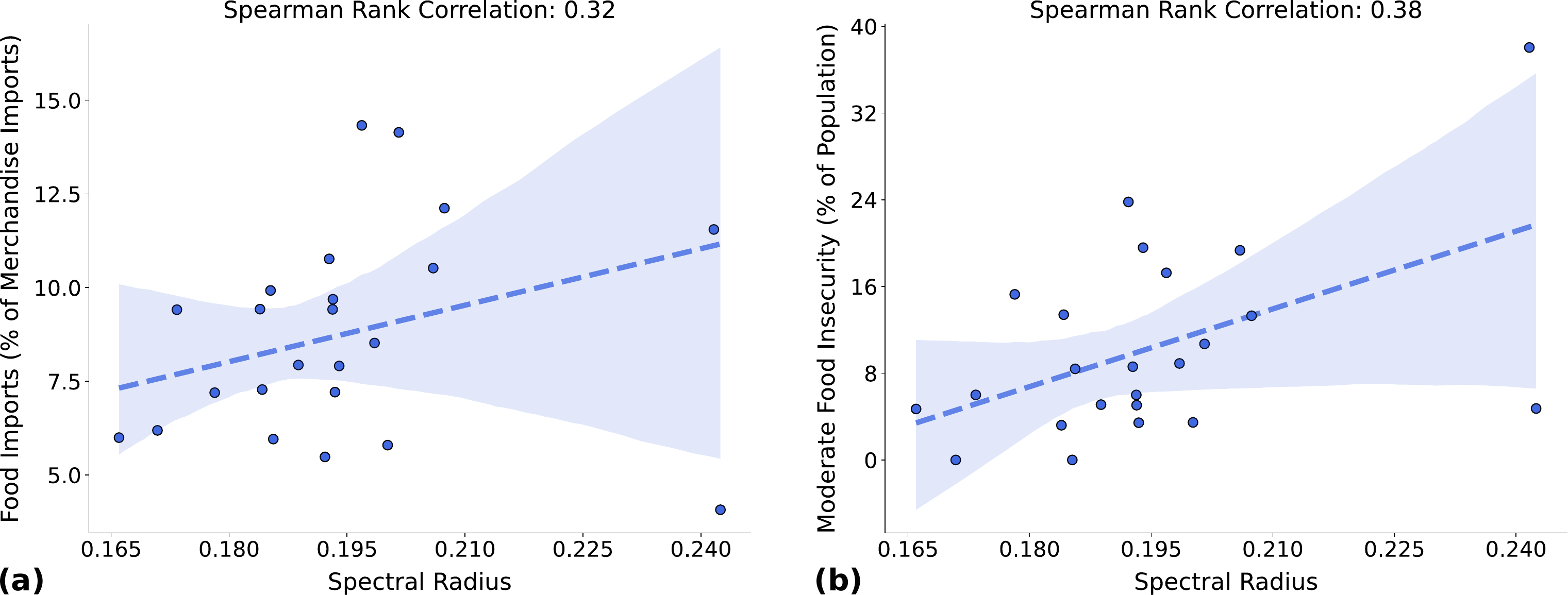}
		\caption[Scatter plot of World Development Indicators vs Spectral Radius for ingredient type graphs]{\textbf{Scatter plot of World Development Indicators vs Spectral Radius for ingredient type graphs}. \textbf{(a)} Relationship between the spectral radius of the ingredient type graphs (x-axis) and the percentage of Food Imports relative to merchandise imports (y-axis), for the geographic regions associated to each cuisine. 
			\textbf{(b)} Relationship between the spectral radius (x-axis) and the percentage of Moderate to Severe Food Insecurity in the population (y-axis), for the same regions.
			Both plots show the best fit line obtained through a robust fit along with the 95\% confidence interval as a blue shaded area. The Spearman rank correlation between the plotted variables is reported at the top of each plot.}
		\label{fig:SI_WDI_spectral_radius}
	\end{figure*}

\begin{figure*}[bh!]
		\centering
		\includegraphics[width=\textwidth]{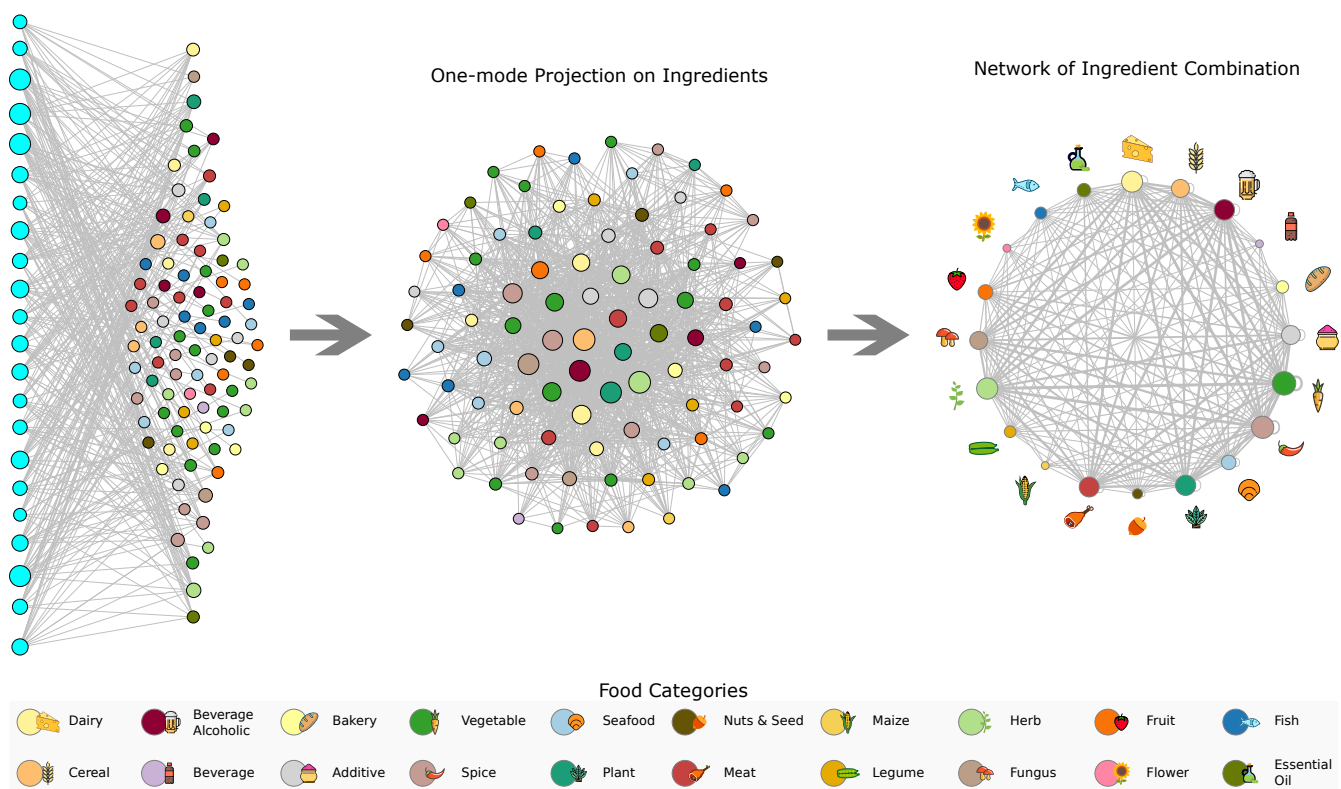}
		\caption[From bipartite representation to network of ingredient combination]{\textbf{From bipartite representation to network of ingredient combination.} The three different network representations of a cuisine (from left to right):  (i) the bipartite recipe-ingredient network, where recipes and ingredients are connected if the ingredient appears in the recipe; (ii) the ingredient projection network, where ingredients are linked if they co-occur in at least one recipe; and (iii) the ingredient type–ingredient type network, where nodes represent ingredient categories and edges capture their co-occurrence across recipes. These representations capture the structure of a cuisine at progressively higher levels of granularity.}
		\label{fig:SI_network_representations}
	\end{figure*}

	We find that both indicators positively correlate with the spectral radius of ingredient type graphs. This relationship is shown in \ref{fig:SI_WDI_spectral_radius}, where each indicator is plotted against the spectral radius.
	
	To analyze these relationships, we performed a robust linear regression fit, which down-weights the influence of outliers in the data. The best-fit line for each relationship is included in the plots. Given the relatively small sample size of our dataset, we use the Spearman Rank Correlation as the measure of correlation. 
	
	{
		\begin{longtable}{p{2.8cm}p{5.3cm}p{2cm}p{6.2cm}}
			\caption{\textbf{Recipes with the Largest Number of Ingredients per Cuisine}} \label{tab:largest_recipes} \\
			\toprule
			\textbf{Cuisine} & \textbf{Recipe Title} & \textbf{\# Ingredients} & \textbf{Ingredient List} \\
			\midrule
			\endfirsthead
			
			\multicolumn{4}{c}%
			{{\tablename\ \thetable{} -- continued from previous page}} \\
			\toprule
			\textbf{Cuisine} & \textbf{Recipe Title} & \textbf{\# Ingredients} & \textbf{Ingredient List} \\
			\midrule
			\endhead
			
			\midrule \multicolumn{4}{r}{{Continued on next page}} \\
			\endfoot
			
			\bottomrule
			\endlastfoot
			African & Morrocan Chili and 10,000 Grains of Sand & 24 &
			apricot; bay leaf; bread; butter; chicken; chili pepper; chive; coriander; cumin; ginger; lamb; lemon; mint; mustard; olive; paprika; parsley; pine; raisin; soy sauce; sugar; tomato; vinegar; water \\[3pt]
			
			American & Turkey Sweet Potato Shepherd's Pie and Cran-applesauce Sundaes & 29 &
			apple sauce; banana; butter; carrot; cayenne; celery; cheese; chicken; cranberry; cream; flour; garlic; ginger; marjoram; mustard; nutmeg; olive; orange; pea; pecan; rosemary; sage; salt; soy sauce; sugar; sweet potato; thyme; turkey; water \\[3pt]
			
			Australasian & Fish Chowder & 21 &
			allspice; bacon; bay laurel; butter; capsicum; cardamom; cayenne; chicken; cinnamon; clam; clove; cod; flour; ginger; mace; milk evaporated; mushroom; mustard; nutmeg; onion; potato \\[3pt]
			
			British Isles & A Scotsman's Shepherd Pie & 21 &
			butter; carrot; cinnamon; cream; egg; flour; garlic; ginger; lamb; mustard; olive; paprika; parsley; pea; potato; salt; soy sauce; sugar; tomato; vinegar; water \\[3pt]
			
			Canadian & Grey Cup Chili & 26 &
			bean; beef; carrot; celery; chili pepper; cilantro; cooking oil; coriander; corn; cumin; egg; garbanzo; garlic; kidney bean; lemon; lime; liquid smoke; mushroom; mustard; onion; onion red; oregano; sugar; vinegar; wheat; zucchini \\[3pt]
			
			Caribbean & Tortilla with Jerk Chicken & 23 &
			allspice; avocado; bean; canola oil; cayenne; cheese; chicken; cinnamon; cream; jalapeno; lime; nutmeg; orange; pepper; rice; sage; salt; scotch; soy sauce; sugar; thyme; vinegar; wheat \\[3pt]
			
			Central European & Venison with Chanterelles in Cream Sauce and Braised Red Cabbage & 20 &
			apple; brandy; bread; butter; caraway; carrot; celery; cream; current red; flour; leek; lemon; meat; mushroom; mustard; rosemary; sherry; soybean sauce; tomato; water \\[3pt]
			
			Chinese & Spicy Hotpot & 20 &
			bean; cabbage; cilantro; cooking oil; corn; ginger; lamb; lemon; liquid smoke; mustard; onion; orange; peanut oil; pepper; shrimp; soybean sauce; star anise; sugar; tomato; vegetable stock \\[3pt]
			
			East European & Bigos (Hunter's Stew) & 22 &
			bacon; basil; bay leaf; beef; cabbage; caraway; carrot; cayenne; cinnamon; flour; ginger; marjoram; mustard; paprika; pork; salt; sauerkraut; soy sauce; sugar; vinegar; water; wine \\[3pt]
			
			French & Porcini Pork Tenderloin & 23 &
			basil; bay laurel; brandy; butter; cream; fennel; garlic; honey; lavender; lemon; marjoram; mushroom; olive; oregano; parsley; pepper; pork; rosemary; shallot; tarragon; thyme; water; winter savory \\[3pt]
			
			Greek & Ground Gyros Platter Deluxe & 22 &
			banana; bread; butter; cheese; chicken; coriander; cumin; flour; ginger; lamb; lemon; mustard; olive; onion; oregano; soy sauce; spinach; sugar; tomato; vinegar; water; yogurt \\[3pt]
			
			Indian & vegetable korma & 31 &
			asafoetida; bay laurel; buttermilk; capsicum; cardamom; cayenne; chili pepper; cinnamon; clove; coriander; cream; cucumber; cumin; curry leaf; french bean; garlic; ghee; ginger; ginger garlic paste; milk; onion; pea; pepper; poppy seed; potato; star anise; sunflower; tomato; turmeric; wheat; zucchini \\[3pt]
			
			Italian & Sauteed Baby Artichokes with Oven-Dried Tomatoes and Green-Olive Dressing & 24 &
			artichoke; basil; bay laurel; chive; cooking oil; cream; dill; egg; fennel; lavender; marjoram; mushroom; olive; onion; oregano; pepper; rosemary; salt; tarragon; thyme; tomato; vegetable broth; wine; winter savory \\[3pt]
			
			Japanese & Homemade Japanese Curry & 22 &
			allspice; apple; cardamom; carrot; chili pepper; clove; coriander; cumin; fennel; fenugreek; flour; mustard; nutmeg; pepper; potato; salt; star anise; sugar; tomato; turmeric; vinegar; water \\[3pt]
			
			Korean & Budae Jjigae (Korean Army Stew) & 19 &
			anchovy; baking powder; butter; chili pepper; cocktail; daikon; egg; fish; flour; kombu; pepper; pork; rice; scallion; sesame; soy sauce; tofu; watercress; wheat \\[3pt]
			
			Mexican & Epi's 50-Ingredient Super Bowl Nachos & 27 &
			artichoke; avocado; bacon; basil; buffalo; celery; chicken; chive; cilantro; cream; garlic; olive; onion red; orange; oregano; paprika; pork; potato; potato chip; salt; scallion; soybean sauce; spinach; tomato; tortilla chip; towel gourd; wheat \\[3pt]
			
			Middle Eastern & Couscous with Lamb Stew & 21 &
			bay leaf; butter; butternut pumpkin; carrot; cayenne; chickpea; chili pepper; habanero; lamb; olive; onion; oregano; paprika; pepper; rose; rosemary; saffron; thyme; turnip; water; zucchini \\[3pt]
			
			Misc. Portuguese & Kale and Cabbage Soup & 22 &
			basil; bay laurel; bay leaf; beef; butter; cabbage; celery; fennel; garlic; kale; kidney bean; lavender; marjoram; olive; oregano; parsley; rosemary; tarragon; thyme; tomato; water; winter savory \\[3pt]
			
			Scandinavian & Mum's Swedish Meatballs & 20 &
			allspice; bread; butter; cinnamon; egg; flour; garlic; ginger; milk evaporated; mustard; nutmeg; olive; parsley; pork; salt; sherry; soy sauce; sugar; vinegar; water \\[3pt]
			
			South American & Peruvian Menestron Soup & 18 &
			basil; bean lima; beef; cabbage; carrot; celery; cheese; corn; garbanzo; garlic; onion; pasta; pea; pepper; potato; spinach; tomato; vegetable oil \\[3pt]
			
			South East Asian & Grilled Steak and Papaya Salad & 25 &
			beef; carrot; chive; cilantro; cooking oil; cream; dill; egg; fish; honey; lemon; lettuce; lime; mint; onion; papaya; parsley; peanut; peanut oil; salt; shallot; soy sauce; sugar; vinegar; watercress \\[3pt]
			
			Spanish & Seafood, Sausage and Bell Pepper Paella & 25 &
			bay leaf; cayenne; coriander; garlic; halibut; ham; lemon; lime; lobster; mussel; olive; onion; oregano; parsley; pea; pork; rice; saffron; salt; shrimp; thyme; tomato; vegetable oil; vinegar; wine \\[3pt]
			
			Thai & Thai Green Curry Chicken & 26 &
			anchovy; bacon; cinnamon; coconut; coconut milk; coriander; cornstarch; cranberry; cumin; curry powder; egg; fish; garlic; jasmine; lemon; lemongrass; lime; olive; onion; peanut; raisin; rosemary; salt; syrup; turmeric; vinegar \\
			\hline
		\end{longtable}
	}
	
	\begin{table}[htbp]
		\centering
		\caption{\textbf{Cuisine-Countries mapping}}
		\label{tab:cusine_country}
		\begin{tabular}{p{3cm}p{13cm}}
			\hline
			\textbf{Cuisine} & \textbf{Countries covered}                                                                                                                   \\ 
			\hline
			Indian           & India, Pakistan, Bangladesh                                                                                                                  \\
			African &
			Nigeria, Ethiopia, DR Congo, Tanzania, South Africa, Kenya, Uganda, Sudan, Algeria, Morocco, Angola, Ghana, Mozambique, Madagascar, Côte d'Ivoire, Cameroon, Niger, Mali, Burkina Faso, Malawi, Zambia, Chad, Somalia, Senegal, Zimbabwe, Guinea, Rwanda, Benin, Burundi, Tunisia, South Sudan, Togo, Sierra Leone, Libya, Congo, Central African Republic, Liberia, Mauritania, Eritrea, Gambia, Botswana, Namibia, Gabon, Lesotho, Guinea-Bissau, Equatorial Guinea, Mauritius, Eswatini, Djibouti, Comoros, Cabo Verde, Sao Tome \& Principe, Seychelles \\
			Chinese          & China, Taiwan                                                                                                                                \\
			Thai             & Thailand                                                                                                                                     \\
			South East Asian & Cambodia, Laos, Vietnam, Philippines, Brunei, Malaysia, Singapore, Indonesia, Timor-Leste                                                    \\
			Middle Eastern   & Yemen, Oman, United Arab Emirates, Kuwait, Saudi Arabia, Jordan, Israel, Lebanon, Syria, Iraq, Iran, Turkiye, Armenia, Egypt, Qatar, Bahrain \\
			Korean           & South Korea, North Korea                                                                                                                     \\
			Japanese         & Japan                                                                                                                                        \\
			Australasian     & Australia, New Zealand                                                                                                                       \\
			Mexican          & Mexico                                                                                                                                       \\
			Caribbean        & Cuba, Jamaica, Haiti, Dominican Republic, Puerto Rico, Trinidad and Tobago                                                                   \\
			South American   & Argentina, Bolivia, Brazil, Chile, Colombia, Ecuador, Guyana, Paraguay, Peru, Suriname, Uruguay, Venezuela                                   \\
			Central European & Germany, Austria, Switzerland                                                                                                                \\
			Greek            & Greece                                                                                                                                       \\
			French           & France                                                                                                                                       \\
			Misc Portuguese  & Portugal                                                                                                                                     \\
			Italian          & Italy                                                                                                                                        \\
			British Isles    & England, Scotland, Wales, Ireland, Isle of Man, Inner and Outer Hebrides, Northern Isles                                                     \\
			Scandinavian     & Norway, Sweden, Finland, Aland Islands, Svalbard                                                                                             \\
			East European    & Poland, Latvia, Lithuania, Belarus, Ukraine, Moldova, Romania, Hungary, Slovakia, Czechia, Bulgaria, Russia                                  \\
			Spanish          & Spain                                                                                                                                        \\
			American         & United States of America                                                                                                                     \\
			Canadian         & Canada                                                                                                                                       \\ 
			\hline
		\end{tabular}
	\end{table}
	
\end{document}